\documentclass[useAMS,usenatbib]{mnras}
\usepackage{amsmath}
\usepackage{appendix}
\usepackage{graphicx}
\usepackage{xfrac}
\usepackage{caption}
\usepackage{array}
\usepackage{longtable}
\usepackage{pdflscape}
\usepackage[figuresright]{rotating}
\usepackage{booktabs}
\usepackage[usenames, dvipsnames]{color}
\usepackage{subcaption}
\usepackage[flushleft]{threeparttable}
\usepackage[export]{adjustbox}
\usepackage{hyperref}
\allowdisplaybreaks

\usepackage{float}

\usepackage{media9}
\usepackage{animate}
\usepackage{etoolbox}

\DeclareRobustCommand{\ion}[2]{%
	\relax\ifmmode
	\ifx\testbx\f@series
	{\mathbf{#1\,\mathsc{#2}}}\else
	{\mathrm{#1\,\mathsc{#2}}}\fi
	\else\textup{#1\,{\mdseries\textsc{#2}}}%
	\fi}

\emergencystretch=\maxdimen
\title[GAMA: Forensic Mass--Metallicity]{Galaxy And Mass Assembly (GAMA): The inferred mass--metallicity relation from $z=$ 0 to 3.5 via forensic SED fitting}
\author[Bellstedt et al.]
{Sabine Bellstedt,$^{1}$\thanks{Email: sabine.bellstedt@uwa.edu.au} Aaron S. G. Robotham,$^{1,2}$ Simon P. Driver,$^{1}$ Jessica E. Thorne,$^{1}$ \newauthor Luke J. M. Davies,$^{1}$ Benne W. Holwerda,$^{3}$ Andrew M. Hopkins,$^{4}$  Maritza A. Lara-Lopez,$^{5}$ \newauthor \'{A}ngel R. L\'{o}pez-S\'{a}nchez,$^{4,6,2}$ Steven Phillipps$^{7}$ 
\\
$^{1}$ ICRAR, The University of Western Australia, 7 Fairway, Crawley WA 6009, Australia\\
$^{2}$ ARC Centre of Excellence for All Sky Astrophysics in 3 Dimensions (ASTRO 3D)  \\
$^{3}$ Department of Physics and Astronomy, University of Louisville, Natural Science 102, Louisville KY 40292, USA \\
$^{4}$ Australian Astronomical Optics, Macquarie University, 105 Delhi Rd, North Ryde, NSW 2113, Australia \\
$^{5}$ Armagh Observatory and Planetarium, College Hill, Armagh, BT61 9DG, UK \\
$^{6}$ Department of Physics and Astronomy, Macquarie University, NSW 2109, Australia\\
$^{7}$ Astrophysics Group, School of Physics, University of Bristol, Bristol BS8 1TL, UK 
}

\begin{document}

\date{}

\pagerange{\pageref{firstpage}--\pageref{lastpage}} \pubyear{2019}

\maketitle

\label{firstpage}

\begin{abstract}

We analyse the metallicity histories of $\sim$4,500 galaxies from the GAMA survey at $z<0.06$ modelled by the SED-fitting code \textsc{ProSpect} using an evolving metallicity implementation. 
These metallicity histories, in combination with the associated star formation histories, allow us to analyse the inferred gas-phase mass--metallicity relation. 
Furthermore, we extract the mass--metallicity relation at a sequence of epochs in cosmic history, to track the evolving mass--metallicity relation with time. 
Through comparison with observations of gas-phase metallicity over a large range of redshifts, we show that, remarkably, our forensic SED analysis has produced an evolving mass--metallicity relationship that is consistent with observations at all epochs. 
We additionally analyse the three dimensional mass--metallicity--SFR space, showing that galaxies occupy a clearly defined plane. This plane is shown to be subtly evolving, displaying an increased tilt with time caused by general enrichment, and also the slowing down of star formation with cosmic time. This evolution is most apparent at lookback times greater than 7 Gyr. The trends in metallicity recovered in this work highlight that the evolving metallicity implementation used within the SED fitting code \textsc{ProSpect} produces reasonable metallicity results over the history of a galaxy. This is expected to provide a significant improvement to the accuracy of the SED fitting outputs. 
\end{abstract}



\section{Introduction}

An analysis of the chemically enriched nature of galaxies, referred to as their metallicity, has become a commonly used tool to study the evolution of galaxy populations. 
The relation between galaxy metallicity and luminosity was readily highlighted in the past \citep{McClure68, Rubin84}, showing that more luminous galaxies tend to have higher metallicities. It has been determined, however, that the scatter in the mass--metallicity is lower than the scatter in the luminosity--metallicity (L-Z) relation \citep{Berg12}, an indication that it is stellar mass that is more fundamentally linked than luminosity to a galaxy's metallicity. The gas-phase mass--metallicity relation (MZR), has since been studied in great detail for dwarf galaxies \citep{Lequeux79, LopezSanchez10c, Berg12, Calabro17, McQuinn20}, large statistical samples of galaxies \citep{Tremonti04, LaraLopez13a, Curti20}, galaxies that fall below the MZR \citep{Peeples09}, and integral field data \citep{Sanchez13, Sanchez19} etc, to understand why this relation exists, and what its implications are for galaxy evolution. 

One of the earliest explanations for this relation came from \citet{Larson74}, who pointed towards gas loss as a means of suppressing metallicity in lower-mass galaxies. 
The gas fraction of a galaxy does seem to impact its position in the MZR, with galaxies deficient in gas tending to have higher metallicities \citep{Hughes13, LaraLopez13b, Zahid14a, Brown18}.
Notably, however, the environment of a galaxy seems to have little or no effect on its position in the MZR, as shown both by \citet{Mouhcine07} with an analysis of large-scale environment. \citet{Hughes13} also investigated the role of the local environment on the MZR, concluding that any environmental trends that could be observed, were likely a second-order effect.

Another key element to interpreting the MZR is understanding how it has evolved with time. Many studies have devoted their attention to the measurement of metallicities in galaxies at high redshifts \citep[including][]{Erb06, Mannucci10, Henry13b, Yabe14, Ly16, Huang19, Weldon20, Sanders20} in order to characterise how the shape and normalisation of the MZR have evolved with time. These studies have shown that the normalisation of the MZR was lower at earlier times, with a similar shape. 

In a study of the stellar populations in spiral galaxies, \citet{Bell00} established that the metallicities of galaxies were dependent on both mass and surface density, where the star formation histories of galaxies were driven by their surface densities. This pointed toward a connection between metallicity and star formation rate (SFR). 
\citet{Ellison08} showed that, at a given stellar mass, the metallicity of a galaxy was higher for galaxies with a lower SFR. 
Analysis of the SFRs of galaxies across the MZR revealed that the MZR was actually a projection of the three dimensional mass--metallicity--SFR relation \citep[for example][]{Mannucci10, LaraLopez10b, Yates12, LaraLopez13a, Brown16, Brown18, Curti20}. We now understand that the status of chemical enrichment of a galaxy is fundamentally linked to the star formation rate, and the build-up of stellar mass.

Observational measurements of gas-phase metallicities in galaxies are typically conducted via measurements of nebular emission lines \citep[for example][]{Savaglio05, Erb06, Yabe14, Huang19, Sanders20}. Depending on which lines are detected for an individual galaxy (typically \ion{O}{iii} and \ion{O}{ii}, but also lines like \ion{N}{ii}, \ion{S}{ii} and \ion{S}{iii}, in addition to H$\alpha$ and H$\beta$), various parameters can be derived which, with the combination of carefully pre-determined calibrations \citep[such as those presented by, for example][]{Pettini04, Kobulnicky04, Bian18} can be converted into a gas-phase metallicity value. 
The impact of both the strong-line parameters and the metallicity calibrations is an extensive field of research, as there are significant systematic differences between the various implementations, as highlighted by \citet{Kewley08} and \citet{LopezSanchez10b}. Consequently, when conducting an analysis of the evolution in the MZR or trends with SFR, stellar mass, gas fractions and environment, the underlying measurement systematics must be carefully considered and accounted for. 

Not only can the signatures of a galaxy's gas-phase metallicity be found in the strength of its nebular emission lines, but the optical range of the SED is also sensitive to variations in the galaxy's gas-phase metallicity. 
In particular, variations in the history of a galaxy's gas-phase metallicity can also influence the SED (as demonstrated by Thorne et al. in prep.).  
With careful modelling, and sufficiently accurate photometric measurements, it is therefore possible to model the metallicity of a galaxy using SED fitting. 
Historically, the metallicity evolution implementation within SED fitting codes has been simplified, with the focus of most methods instead being placed on the parametrisation of a galaxy's star formation history. 
As shown in recent work that uses the SED fitting code \textsc{ProSpect} \citep{Robotham20} to recover the cosmic star formation history \citep{Bellstedt20b}, there are significant benefits to be gained when carefully modelling the evolving gas-phase metallicity in galaxies, rather than simply assuming this value to be constant over time. 

In this paper, we extend the work presented by \citet{Bellstedt20b} to show that not only can this technique accurately reproduce the star formation history of the Universe, but it can also successfully recover the metallicity distributions of galaxy populations. 
This recovery is important, as it highlights that a more complex approach to metallicity modelling in SED fitting can produce physical results in a broad range of parameter spaces. 
We describe the GAMA data in Sec. \ref{sec:Data}, and the SED fitting technique in Sec. \ref{sec:ProSpectFitting}. Our derived MZR is presented in Sec. \ref{sec:MZR}, followed by an analysis of the evolving MZR in Sec. \ref{sec:EvolvingMZR}. We additionally present our derivation of the mass--metallicity--SFR plane evolving through cosmic time in Sec. \ref{sec:Plane}. We finally discuss these results in Sec. \ref{sec:Discussion}. 

For all stellar mass measurements presented in this work, we have utilised a \citet{Chabrier03} initial mass function (IMF). The cosmology assumed throughout this paper is $H_0 = 67.8\,\rm{km}\,\rm{s}^{-1}\,\rm{Mpc}^{-1}$,  $\Omega_m = 0.308$ and $\Omega_{\Lambda} = 0.692$ \citep[consistent with a Planck 15 cosmology][]{Planck16}. 


\section{Data}
\label{sec:Data}

The GAMA survey \citep{Driver11, Liske15} is a large program that gathered redshifts for $\sim$300,000 galaxies across five different fields spanning 230 square degrees using the Anglo Australian Telescope. One of the great strengths of the survey is that it achieved a high spectroscopic completeness of 98\% above a magnitude limit of $m_r \leq 19.5$\footnote{With the development of the updated photometric catalogue for the survey \citep{Bellstedt20}, the completeness limit has been updated from $m_r \leq 19.8$ to $m_r \leq 19.5$ in the equatorial fields. } (or $m_i \leq 19.0$ in the G23 field). 

In this work, as in \citet{Bellstedt20b}, we utilise the spectroscopic and photometric data for 6,688 galaxies with $z<0.06$ and $m_r \leq 19.5$ in the three equatorial fields (G09, G12, G15) to conduct SED fitting using the code \textsc{ProSpect} \citep{Robotham20}. The panchromatic photometry catalogue for the GAMA survey was recently updated to include the KiDS imaging in the optical bands, and also to utilise the \textsc{ProFound} source detection software \citep{Robotham18}. This updated photometry was presented in \citet{Bellstedt20}. 
These data include photometry in 19 bands from the far-UV (FUV) to the far-IR (FIR). 

As in \citet{Bellstedt20b}, we do not explicitly identify potential AGN in the sample to be removed. We expect that this will have a minimal impact on our results, as the AGN contamination in this sample is expected to be very small (fewer than 30 galaxies, \citealt{Prescott16}). 

In this work we use \texttt{v1} of the \texttt{GAMAKidsVikingFIR} DMU.


\section{SED fitting}
\label{sec:ProSpectFitting}

We implement the same method as outlined in \citet{Bellstedt20b}, using the GAMA photometry presented in \citet{Bellstedt20} and passed into the \textsc{ProSpect} SED fitting code \citep{Robotham20}. 
For a detailed description of the fitting we direct the reader to \citet{Bellstedt20b}, however we provide a brief summary in this section. 
The stellar templates we use in this analysis are from \citet{Bruzual03}, and the star formation history is parametrised by the \texttt{massfunc\_snorm\_trunc} function within \textsc{ProSpect}. 
This parametrisation takes on the form of a skewed Normal distribution, with the peak position (\texttt{mpeak}), peak SFR (\texttt{mSFR}), SFH width (\texttt{mperiod}), and SFH skewness (\texttt{mskew}) set as free parameters. 
A positive value of the skewness produces a SFH tailing off toward the present day, whereas a negative skewness causes the SFH to tail off towards the start of the Universe. 
The SFH is anchored to $0$ at a lookback time of 13.4 Gyr, deemed in this work to be the age at which galaxies start forming. As outlined by \citet{Bellstedt20b}, this value was selected to correspond with the epoch at which the highest-$z$ galaxies are known to exist ($z=11$, \citealt{Oesch16}).

In this \textsc{ProSpect} analysis, we implement an evolving metallicity where the shape of the stellar mass evolution is linearly mapped onto the shape of the metallicity evolution for each galaxy, given by the \texttt{Zfunc\_massmap\_lin} function. This ensures that chemical enrichment in the galaxies follows the assumed star formation rate, where increased star formation is associated with an increased rate of metal production in the galaxy. 
The final metallicity of the galaxy is allowed to be a free parameter, \texttt{Zfinal}. We highlight that this value represents the present-day gas-phase metallicity of the object, and correspondingly the metallicity of the youngest stars in the galaxy (as opposed to a time-averaged stellar metallicity). 
This approach is a significant improvement over the typical approach in SED fitting, which is to assume that the metallicity is constant over a galaxy's history. The impact of the metallicity assumption on the cosmic star formation history (CSFH) was demonstrated in fig. 4 of \citet{Bellstedt20b}. 
The range in \texttt{Zfinal} values is limited by the range of metallicity in the \citet{Bruzual03} stellar templates. The upper limit of these templates is 0.05, and hence our recovered \texttt{Zfinal} values cannot extend beyond this value. The resulting CSFH and SMD that are derived using this metallicity implementation are shown in appendix B of \citet{Bellstedt20b}. 

The fitting outputs used in this work were first presented by \citet{Bellstedt20b}, where they were used to derive the cosmic star formation history, and the cosmic metal density evolution. While the main results in that study were derived using a closed-box metallicity implementation in \textsc{ProSpect} (as given by the \texttt{Zfunc\_massmap\_box} function), appendix B presented the cosmic star formation history and cosmic stellar mass density when assuming linear metallicity evolution, as given by the \texttt{Zfunc\_massmap\_lin} function. Because the yields are not assumed to be constant in the linear metallicity implementation (unlike the closed-box implementation), the late-time enrichment of galaxies is slightly reduced when this metallicity evolution is prescribed. 
While the selected metallicity implementation does not have a large impact on the results, we noted that the number of objects hitting the upper metallicity limit is lower when assuming linear metallicity evolution. We interpret this to indicate that the metallicity outputs derived when using the \texttt{Zfunc\_massmap\_lin} are more physical. 
This assumption of allowing the metallicity to grow in proportion to the stellar mass growth is similar to what is seen in the chemical enrichment of galaxies in semi-analytic models \citep[as seen in, for example,][]{Robotham20}, providing motivation to use this implementation. 
See the discussion in Appendix \ref{sec:LinearityValidity} for a more detailed analysis of the enrichment in \textsc{Shark}, highlighting the degree to which the proportionality assumption is accurate in this semi-analtyic model.
As such, we use the outputs as derived by the \texttt{Zfunc\_massmap\_lin} function in this work. 

In addition to the five free parameters specifying the star formation and metallicity histories, we include four free parameters to describe the dust contribution to the SED. The dust is assumed to exist in two forms; either in birth clouds formed around young stars, or distributed as a screen in the interstellar medium (ISM). For each of these components, we include two free parameters, describing the dust opacity (\texttt{tau\_birth}, \texttt{tau\_screen}), and the dust radiation field intensity (\texttt{alpha\_birth}, \texttt{alpha\_screen}). 
Hence, we model the SED in our work using a total of nine free parameters. The fitting ranges and priors are presented in table 2 of \citet{Bellstedt20b}. 

Out of the 6,688 galaxies analysed in the $z<0.06$ sample presented in \citet{Bellstedt20b}, in this analysis we focus only on a subset of these objects that have a reasonable constraint on the metallicity parameter from SED fitting. In order to determine this subset, we remove from analysis any objects for which the $1\sigma$ uncertainty from the MCMC sampling is greater than 0.5 dex for the \texttt{Zfinal} parameter. After removing these objects, we are left with a sample of 4,531 galaxies for which we have constrained metallicity estimates. 

\subsection{Comparison to SDSS metallicities}

\begin{figure}
	\centering
	\includegraphics[width=90mm]{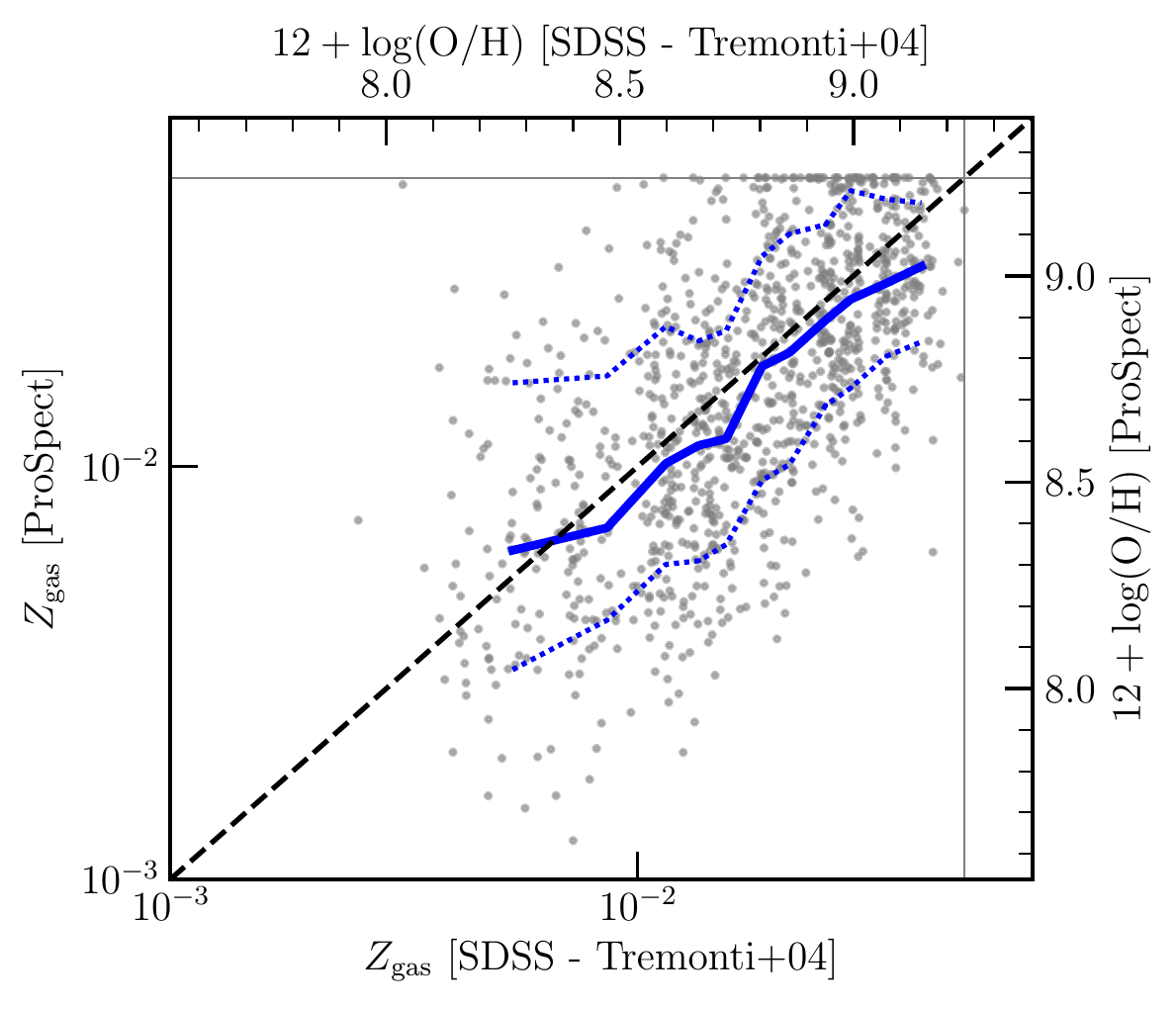}
	\caption{Comparison of the \textsc{ProSpect}-derived metallicity values with the corresponding measured metallicites from \citet{Tremonti04} for a matched sample. The dashed black line shows the one-to-one, and the thin grey vertical and horizontal lines show the upper metallicity limit of the \citet{Bruzual03} templates. The solid blue line shows the running median, with the 1$\sigma$ scatter indicated by the dotted blue lines. The scatter in our metallicity recovery is $\sim$ 0.25 dex, with an offset of $-0.06$ dex. }
	\label{fig:MetallicityComparison}
\end{figure}

A subset of the objects analysed in this work have spectroscopically-derived metallicities from SDSS\footnote{\url{https://wwwmpa.mpa-garching.mpg.de/SDSS/DR7/oh.html}} \citep{Tremonti04, Brinchmann04}. 
We compare our derived metallicity values against these observational measurements as an assessment of the accuracy of our SED-fitting approach to modelling this quantity. 
The subset consists of 2,220 objects, and a comparison of the metallicities is presented in Fig. \ref{fig:MetallicityComparison}. 
Here we show that the SED-derived values follow the spectroscopically-derived values generally well, with a mean offset of $-0.06$ dex and a scatter of $~\sim$ 0.25 dex. 
Interestingly those values in our sample that are hitting the upper metallicity limit as governed by the upper limit of the \citet{Bruzual03} templates cover a range of values as determined by \citet{Tremonti04}.
Although the \citet{Tremonti04} measurements are not restricted to the same upper limit as the \citet{Bruzual03} templates, the highest metallicities are similar to this limit. 

The SED-derived uncertainties are significantly larger than the spectroscopically-derived uncertainties, which is indicative of the lower constraint that broadband photometry is able to provide. 
The broad agreement between the observationally-measured values from SDSS and the inferred values from our SED fitting highlight that the values we recover are not simply ``nuisance" parameters, and are instead physically meaningful (albeit with significant uncertainty).

\section{The Mass--Metallicity Relation}
\label{sec:MZR}

\begin{figure}
	\centering
	\includegraphics[width=90mm]{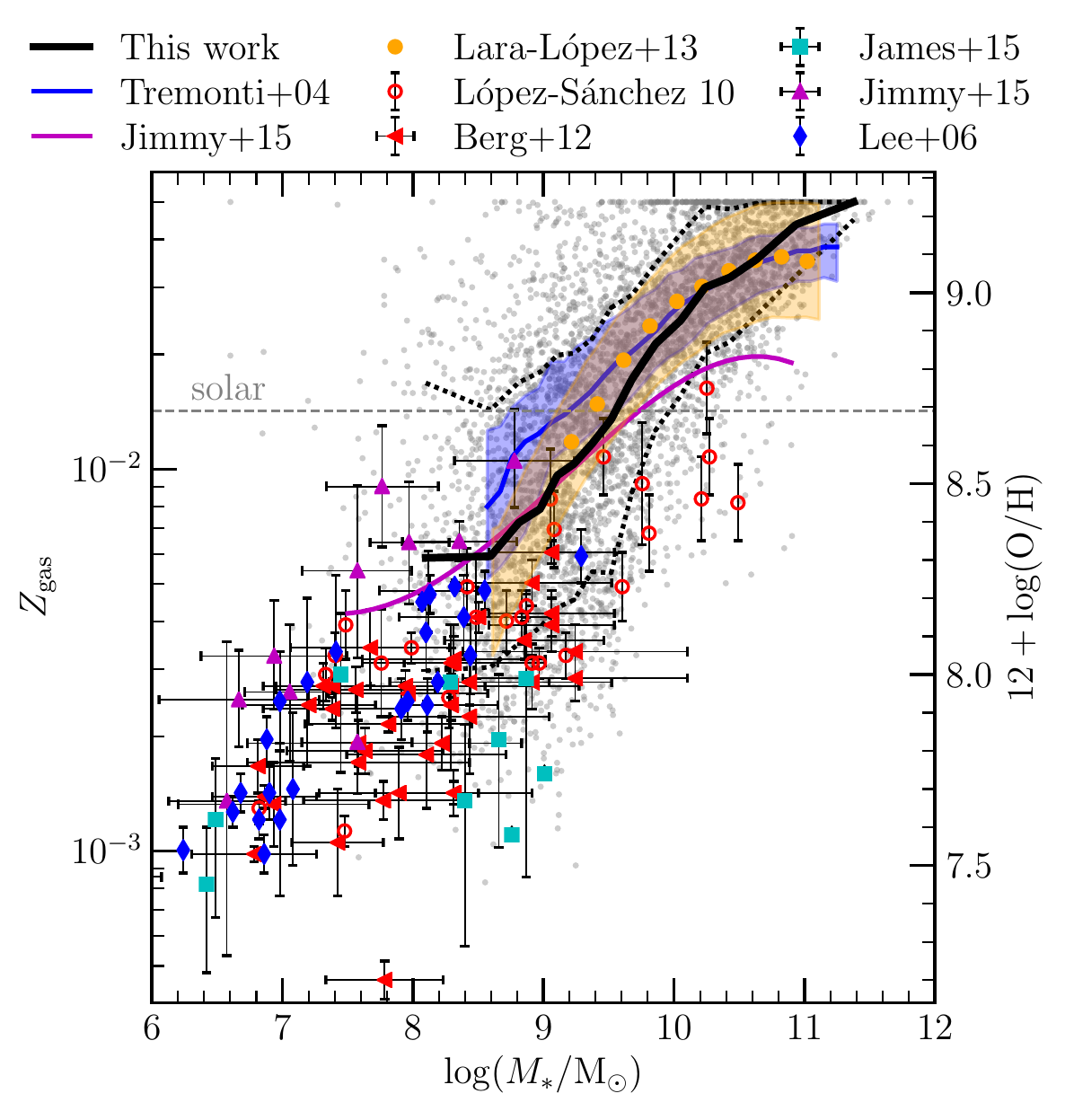}
	\caption{The resulting mass-metallicity relationship when assuming a metallicity that evolves with the star formation history. The blue line indicates the median MZR recovered by \citet{Tremonti04}, orange points indicate the binned metallicity measurements by \citet{LaraLopez13c}, and the magenta line shows the fit to the MZR by \citet{Jimmy15}. We additionally include observations that extend to lower stellar masses from \citet{Lee06, LopezSanchez10c, Berg12, James15, Jimmy15}.  }
	\label{fig:MassMetallicity}
\end{figure}

While there was no fitting prior set on the resulting gas-phase metallicity values in our implementation, we recover in our analysis a mass--metallicity distribution that is consistent with trends recovered by other observations. 
This is shown in Fig. \ref{fig:MassMetallicity}, where for each galaxy in our sample, we plot the resulting stellar mass against the fitted \texttt{Zfinal} value. 
The overall trend of our MZR is shown via the solid black line showing the moving median, and the dashed black lines that indicate the $1\sigma$ range in metallicity at any given stellar mass. In calculating these values, we demand that each bin includes at least 300 galaxies. 
The scatter in the relation at stellar masses below $10^{10}\,\rm{M}_{\odot}$ is significant, but this scatter reduces at larger stellar masses. We see a bending of this relation at $M_*\sim10^{10}\,\rm{M}_{\odot}$. 
Below $M_*\sim10^{9}\,\rm{M}_{\odot}$ our sample becomes increasingly incomplete, so the MZR at these masses is prone to bias. 

A clear artefact in this image is the upper limit in the range of metallicity values at 0.05. 
This is particuarly stark at $M_*>10^{10.5} \,\rm{M}_{\odot}$, where the upper region of our $1\sigma$ range is at this limit. 
This limit is the highest-metallicity template present in the \citet{Bruzual03} stellar population templates, and hence our application of \textsc{ProSpect} is not sensitive to gas-phase metallicity values larger than $Z_{\rm gas} = 0.05$. 

To compare against observational trends, we include measurements of metallicity made for GAMA galaxies by \citet{LaraLopez13c}, and the MZR fit with one sigma scatter produced using SDSS galaxies by \citet{Tremonti04}. 
We additionally include the fit to the MZR presented by \citet{Jimmy15}, in which more massive galaxies were determined using the ALFALFA/SDSS sample. 
Note that this MZR is significantly offset to both the relation recovered by \textsc{ProSpect}, and also to the observational measurements of \citet{Tremonti04} and \citet{LaraLopez13c}. 
We summarise the parameters and calibrators used by each of these studies to determine metallicities in Table \ref{tab:Indicators}. 

\begin{table*}
	\centering
	\caption[Indicators]{Metallicity indicators and calibrations applied in each of the observational studies presented in Fig. \ref{fig:MassMetallicity}. T$_e$ refers to the electron temperature, which is derived using auroral lines, and is regarded as a ``direct" method of measuring metallicity. PP04 refers to \citet{Pettini04}, and CL01 refers to \citet{Charlot01}.  }
	\label{tab:Indicators}
	\begin{tabular}{@{}c | c c | p{4cm} }
		\hline
		Study & Parameters/Emission lines & Calibration  & Comments \\
		\hline
		\citet{Tremonti04} & [\ion{O}{ii}], H$\beta$, [\ion{O}{iii}], H$\alpha$, [\ion{N}{ii}], [\ion{S}{ii}]  &  & Simultaneous fit using CL01\\
		\citet{Lee06} & [\ion{O}{iii}]$\lambda$4363 &  & T$_e$ derivation\\
		\citet{LopezSanchez10c} & [\ion{O}{iii}]($\lambda$4959+$\lambda$5507)/$\lambda$4363, [\ion{N}{ii}]($\lambda$6548+$\lambda$6583)/$\lambda$5755,  &  & T$_e$ derivation\\
		 &  [\ion{O}{ii}]($\lambda$3727+$\lambda$3729)/($\lambda$7319+$\lambda$7330) &  & \\
		 \citet{LaraLopez13c} & O3N2 & PP04 &Strong line derivation\\
		\citet{Berg12} &[\ion{O}{iii}]$\lambda$4363 &  & T$_e$ derivation\\
		\citet{Jimmy15} & N2 & \citet{Denicolo02} & Strong line derivation\\
		\citet{James15} & [\ion{O}{ii}]$\lambda$3727,$\lambda$3729  &  &T$_e$ derivation\\
		\hline		
	\end{tabular}
\end{table*}

While the aforementioned MZR measurements are made using massive galaxies, we also include in Fig. \ref{fig:MassMetallicity} a comparison to metallicity measurements made for galaxies in the dwarf regime by \citet{Lee06}, \citet{LopezSanchez10c}\footnote{We exclude from Fig. \ref{fig:MassMetallicity} the galaxies from \citet{LopezSanchez10c} that are undergoing mergers.}, \citet{Berg12}, \citet{James15}, \citet{Jimmy15}, in each case correcting the stellar masses to a Chabrier IMF where necessary using the conversion factors presented in table 1 of \citet{Driver13}. 
We find that, while these observations overlap with the distribution of points derived by \textsc{ProSpect}, these metallicity measurements (all derived using T$_{e}$ methods) all have on average slightly lower values than those we derive in the low-mass range. 
Similarly, these values are also systematically lower than the observations by \citet{Tremonti04}, \citet{LaraLopez13c} and \citet{Jimmy15} for the overlapping mass range. 

\begin{figure}
	\centering
	\includegraphics[width=90mm]{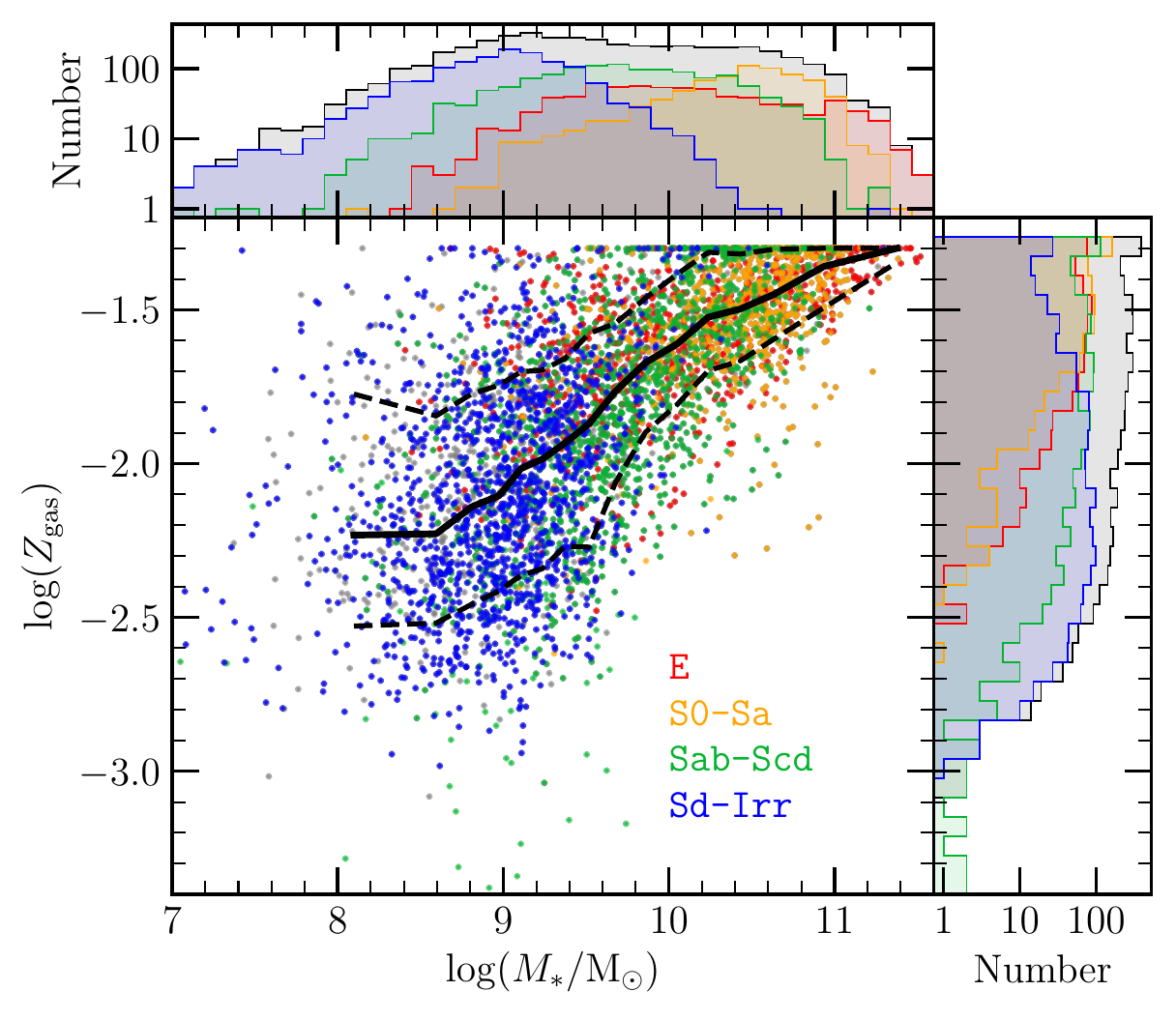}
	\caption{The mass--metallicity relation, as divided by the visually classified morphological types.  }
	\label{fig:MassMetallicity_Morph}
\end{figure}

At the highest stellar masses ($M_*>10^{10.5} \,\rm{M}_{\odot}$) our median MZR is greater than that measured by other observations. 
Observations by \citet{Tremonti04} and \citet{LaraLopez13c} are necessarily restricted to galaxies with star-formation, as the metallicity measurement is made on the emission lines that are produced by star formation. 
Early-type galaxies like ellipticals and lenticulars are very metal-rich and dominate the massive end of the MZR, and as such the high-mass MZR may be biased high in our analysis. 
We determine that a cut in specific SFR does not cause the median metallicity value at high-mass to reduce, however, and therefore the presence of early-type galaxies in our analysis is unlikely to account for the larger metallicities that we derive at large stellar masses.

We show how galaxies with different visual morphologies contribute to the MZR in Fig. \ref{fig:MassMetallicity_Morph}. 
Here, elliptical galaxies are shown in red, \texttt{S0-Sa} galaxies in orange, \texttt{Sab-Scd} galaxies in green, and \texttt{Sd-Irr} galaxies in blue. 
The histogram above the main panel of the plot shows how the morphologies are distributed with stellar mass, whereas the histogram to the right of the main panel shows how they are distributed with gas-phase metallicity. 
Fig. \ref{fig:MassMetallicity_Morph} highlights that early-type galaxies dominate the MZR at high mass and high metallicity, whereas the low metallicity portion of the plot is almost entirely occupied by late-type galaxies. 
Additionally, Fig. \ref{fig:MassMetallicity_Morph} shows that the dispersion of the MZR in the low-metallicity regime (dominated by \texttt{Sd-Irr} galaxies) is much higher than the high-mass regime.

\subsection{Comparison with simulations}

The MZR has historically been very difficult for simulations and semi-analytic models (SAMs) to reproduce, due to the intricate nature of chemical evolution in galaxies. 

\begin{figure}
	\centering
	\includegraphics[width=90mm]{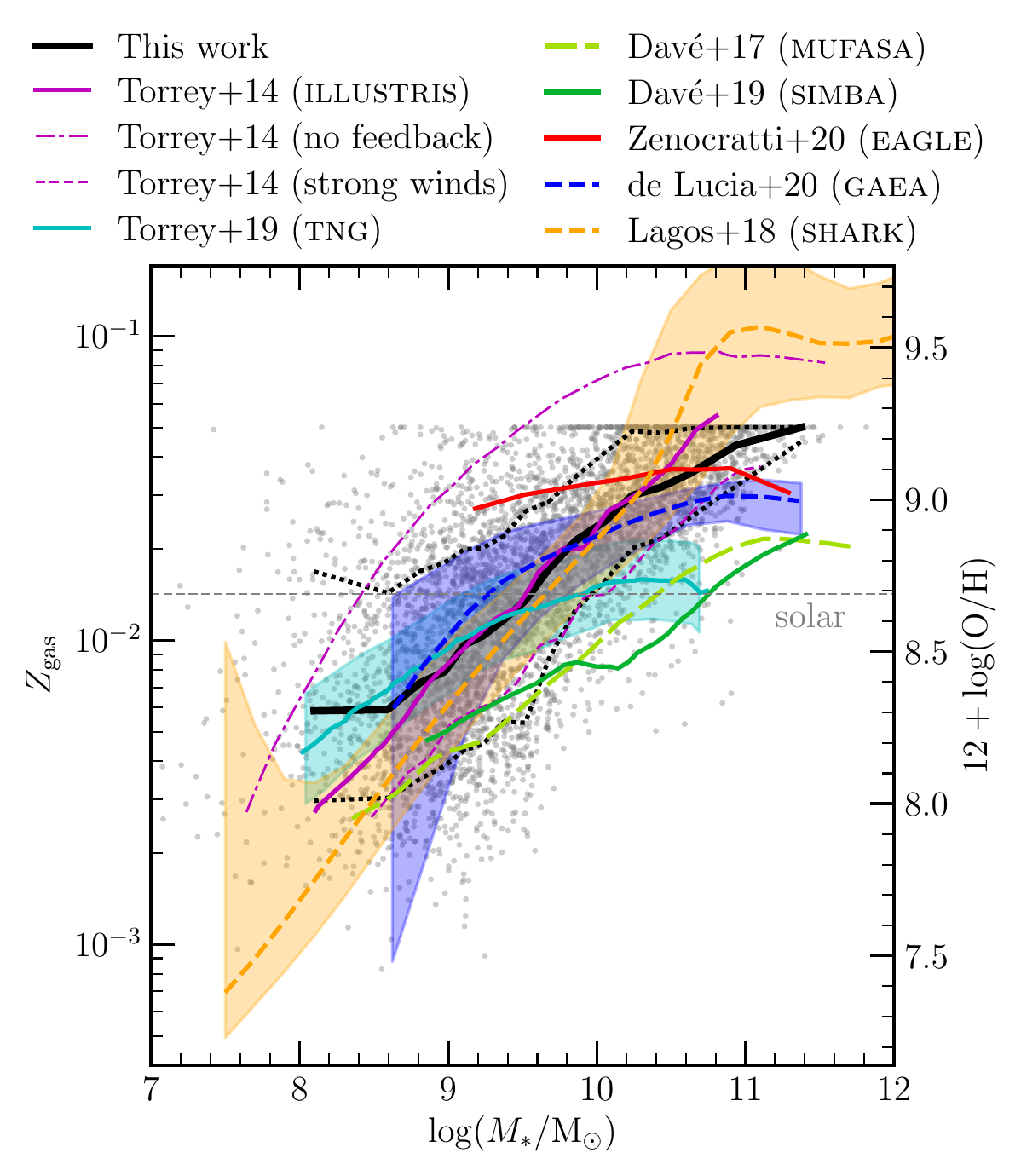}
	\caption{Comparison of our MZR with simulations. Here we include the cosmological hydrodynamic simulations Illustris \citep[][both the default model, and also variations of the feedback model including no feedback, and string winds]{Torrey14}, IllustrisTNG \citep{Torrey19}, MUFASA \citep{Dave17}, SIMBA \citep{Dave19}, EAGLE \citep{Zenocratti20}, and the semi-analytic models GAEA \citep{deLucia20} and \textsc{Shark} \citep{Lagos18}.   }
	\label{fig:MassMetallicity_sims}
\end{figure}

In recent years, there has been an increased reporting of studies that are producing MZR trends more like observations. 
In Fig. \ref{fig:MassMetallicity_sims} we briefly compare our derived MZR with those prodocued by leading simulations/SAMs.
We include the MZR derived by the cosmological, hydrodynamic simulations Illustris \citep{Torrey14}, IllustrisTNG \citep{Torrey19}, MUFASA \citep{Dave17}, SIMBA \citep{Dave19} and EAGLE \citep{Zenocratti20}, and the semi-analytic models GAEA \citep{deLucia20} and \textsc{Shark} \citep{Lagos18}.

\citet{Torrey14} demonstrate with Illustris that the adopted feedback model has a dramatic influence on the resulting MZR, with no feedback resulting in much higher metallicities (shown in Fig. \ref{fig:MassMetallicity_sims} by the dashed magenta line that extends beyond the upper limit of our probed metallicity range), whilst strong feedback reduces the normalisation of the MZR. 
While the bending of the MZR was recovered by Illustris when feedback was removed, the bending of the MZR at high stellar masses was not present when feedback was included. 
For the default feedback model however, the agreement between the Illustris MZR and the MZR recovered in this work is consistent at stellar masses $M_* < 10^{10.5}\rm{M}_{\odot}$. 

Unlike Illustris, where little bending was observed, the MZR for IllustrisTNG \citep[][shown in Fig. \ref{fig:MassMetallicity_sims} in cyan]{Torrey19} does recover a saturation metallicity. 
Both the stellar mass at which this bending occurs, as well as the metallicity value, is significantly lower in IllustrisTNG than we recover using \textsc{ProSpect}. 

In addition to a systematically lower normalisation, the shape of the MZR recovered by SIMBA \citep[][shown in green]{Dave19} is different to that of other simulations. 
It displays a dip at around $10^{10}\rm{M}_{\odot}$, without an indication of a saturation in metallicity at the highest stellar masses. Interestingly a predecessor of SIMBA, MUFASA \citep[][dashed light green]{Dave17}, displays an MZR shape that is more consistent with the other trends presented in Fig. \ref{fig:MassMetallicity_sims}. 
This is despite the fact that \citet{Dave19} describe the MUFASA MZR as being too steep. 

We also include in Fig. \ref{fig:MassMetallicity_sims} the MZR from the EAGLE simulations by \citet[][solid red line]{Zenocratti20}. 
On average, EAGLE galaxies seem to have a higher metallicity than other simulations and observations, with supersolar values at all stellar masses. 
Furthermore, EAGLE does not recover the characteristic MZR shape with lower metallicities in low-mass galaxies, displaying instead a relatively constant metallicity with varying stellar mass. 

The GAEA MZR \citep[][dashed blue line]{deLucia20} recovers the typical lower metallicities for low-mass galaxies, and a saturation of metallicities for high-mass galaxies, although the difference in metallicity between low-mass and high-mass systems is less extreme than what we derive. 
The scatter from \citet{deLucia20} was reported to be much larger than observations at lower stellar masses, although we note that their scatter is only slightly larger than the $1\sigma$ range that we derive. 
The MZR from the semi-analytic model \textsc{Shark} \citep[][dashed orange line]{Lagos18} is consistent in the stellar mass range $10^{9} < M_*/\rm{M}_{\odot} < 10^{10.5}$, however at the high-mass end the predicted MZR is significantly higher than what we infer. 

Simulations frequently compare against observations of the MZR as a way of assessing how closely the physical implementation of the simulation can reproduce reality. 
One set of observations that are frequently used for comparison is the SDSS dataset by \citet{Tremonti04}, shown in Figures \ref{fig:MetallicityComparison} and \ref{fig:MassMetallicity}. 
Due to the biases introduced by different methods of metallicity calibration \citep[as emphasised in the work by ][]{Kewley08}, the absolute normalisation of the \citet{Tremonti04} dataset is debated in the literature. 
When compared against simulations in the works by \citet{Torrey19}, \citet{Dave17} and \citet{Dave19}, the \citet{Tremonti04} vaues were scaled downwards by 0.26 dex, to be consistent with the calibration of \citet{Pettini04}. 
As a result, the agreement between the observations and simulations in those works appears closer than the agreement presented in our Fig. \ref{fig:MassMetallicity_sims}. 
The complexity of comparing observational metallicity measurements with simulations is discussed in detail in \citet{Lagos12}.

In all of the mass--metallicity relations recovered by simulations, the lower mass limit in each comparison is determined by the resolution limit of the underlying dark matter in each simulation. 
As a result, the simulated behaviour at low stellar masses cannot be determined. 
As such, there is no way of comparing whether the low-mass turnover in the MZR we recover is also replicated by either hydrodynamic simulations or semi-analytic models. 

The important aspect to consider when comparing different simulations, is that current state-of-the-art models (either cosmological or semi-analytic) of galaxy evolution still predictly widely varying mass-metallicity relations. 
In fact, the variation between these MZR predictions is significantly greater than the discrepancies we see between our inferred MZR and that of observations. 
This highlights that our approach to metallicity evolution in galaxies is competitive, if not perfect.

\subsection{Evolution in the MZR}
\label{sec:EvolvingMZR}

\begin{figure*}
	\centering
	\includegraphics[width=180mm]{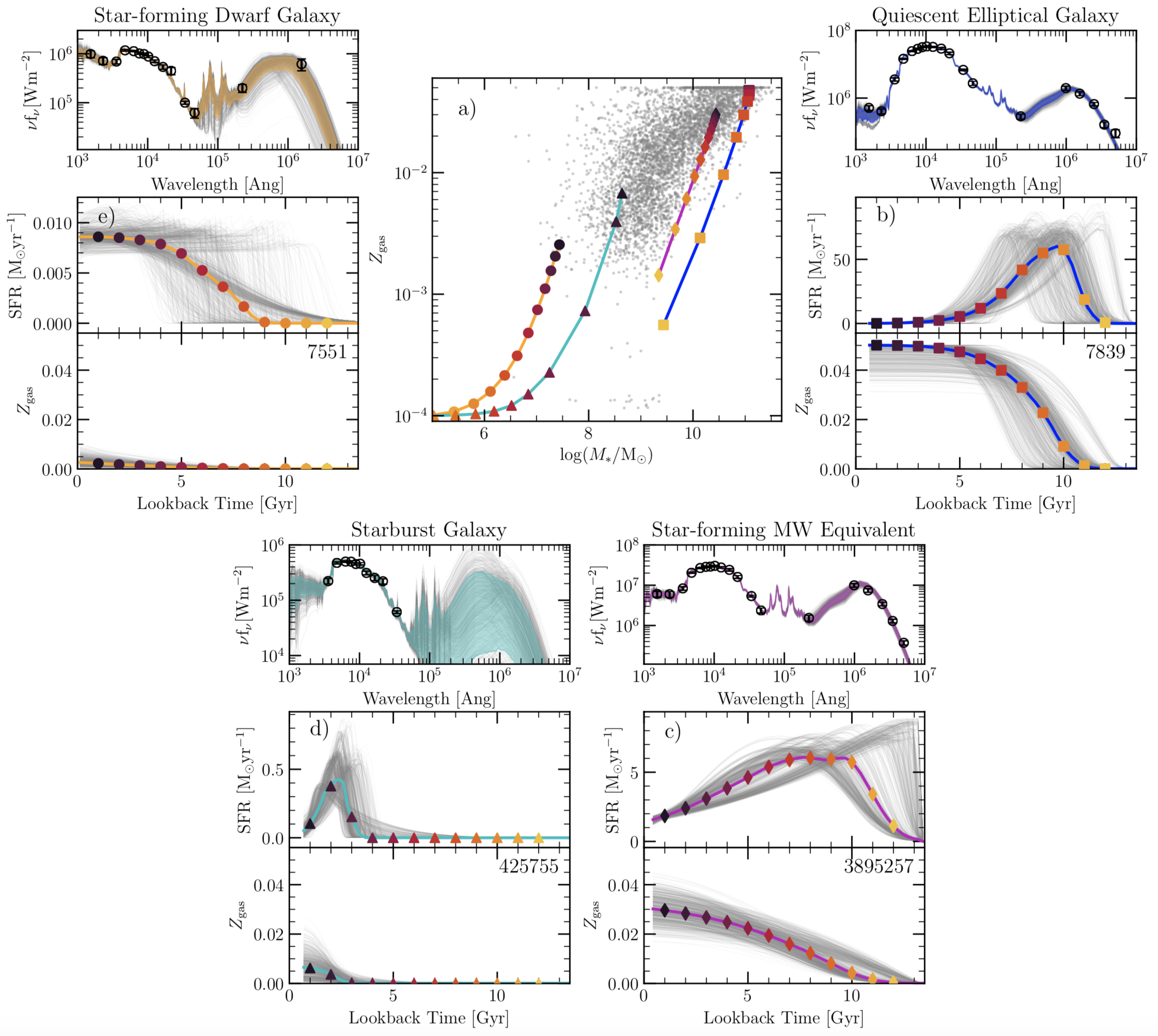}
	\caption{Tracks of four individual galaxies in the mass--metallicity plot a) over cosmic time. For each example galaxy, we show the modelled star formation histories and metallicity histories, as well as the model fit to the SED. These galaxies are: b) 7839; c) 3895257; d) 425755; and e) 7551. The SFR and metallicity values at 1 Gyr intervals are shown in coloured points, with the corresponding position on the mass--metallicity plot shown in panel a). Symbols are used to differentiate between different example galaxies. The distribution of mass--metallicity points for our full sample is shown as grey points in panel a).   }
	\label{fig:MassMetallicityIndividual}
\end{figure*}

Due to the forensic nature of our analysis, we can not only extract the $z=0$ mass-metallicity relation, but also the evolution of this relation over cosmic time, by analysing the star formation and metallicity histories derived using \textsc{ProSpect} for individual galaxies. 
As such, at any arbitrary value of lookback time, we can determine the forensically-inferred gas-phase metallicity, SFR and stellar mass at that epoch. 
This enables us to contruct the MZR of the sample not only at the epoch of observation, but also at other epochs in the Universe's history. 
The following sections will present an analysis of the evolving MZR that we derive. 

The manner in which we use our metallicity and star formation histories to trace back the MZR is shown in Fig. \ref{fig:MassMetallicityIndividual}. 
For four example galaxies (7551, 425755, 3895257 and 7839), we present the star formation history (SFH) and metallicity history (ZH) as modelled by \textsc{ProSpect}, in a two-panel subplot, with the corresponding model fit to the SED shown above. 
The MCMC sampled distribution is shown as 1000 grey lines, and the median SFH and ZH are shown as solid coloured lines. For each of these histories, we indicate the value at 1 Gyr intervals using coloured points. 
In the top middle panel, we show the mass--metallicity plot at $z=0$, where a track has been added for each example galaxy. 
These example galaxies vary from low-mass galaxies that form their stars more recently, to massive galaxies that formed their stars early in the Universe. 
This is evident in the mass--metallicity plot, where each galaxy has a track in a different part of the parameter space. 
It is interesting to note that these tracks bear much resemblance to the various tracks for galaxies of different morphologies presented in fig. 5 by \citet{Calura09}, who produced theoretical chemical enrichment models of galaxies with different morphologies by applying assumptions about chemical enrichment via star formation, inflows and outflows. 

\subsubsection{Comparison with observations}

\begin{figure*}
	\centering
	\includegraphics[width=180mm]{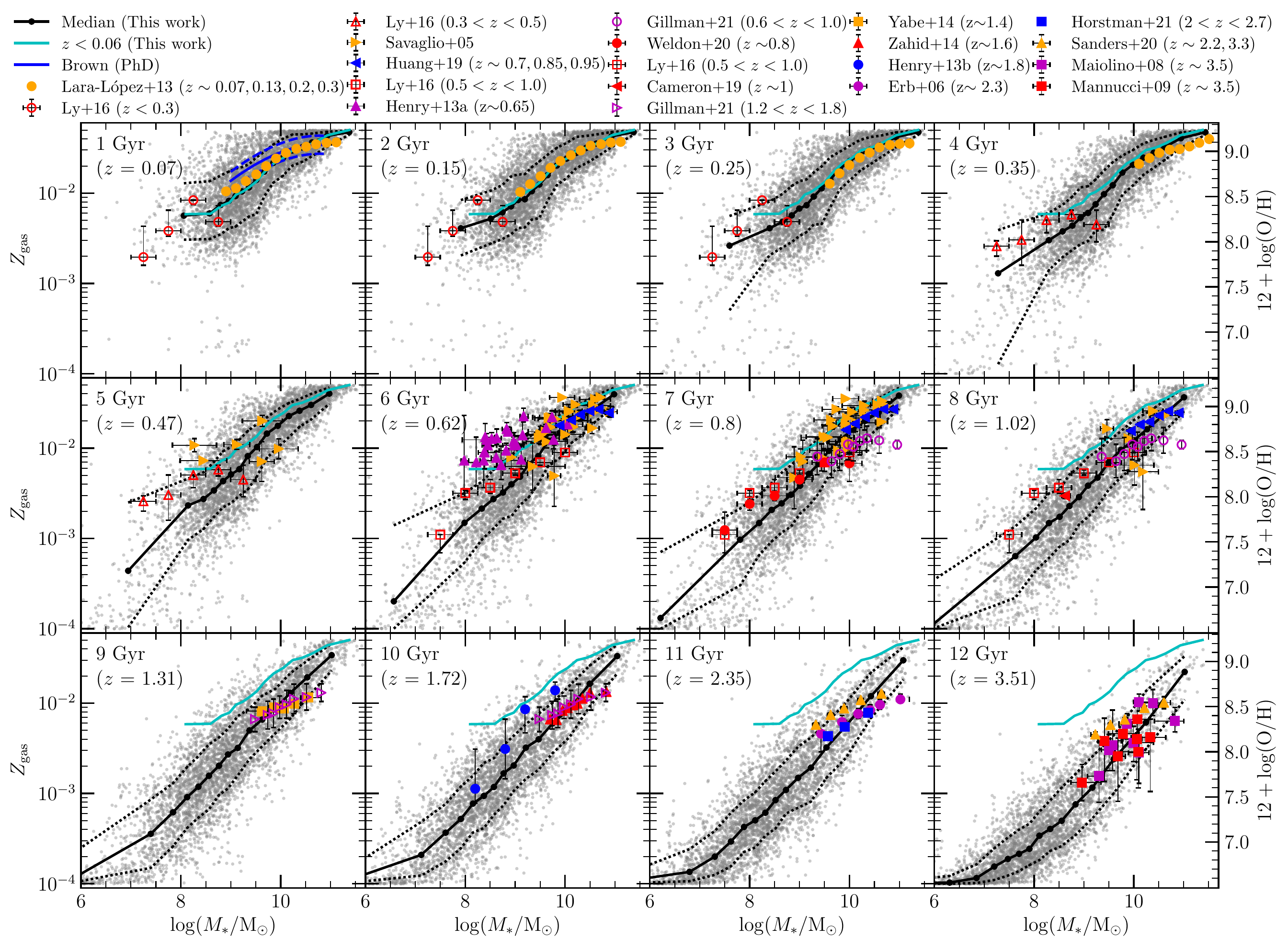}
	\caption{The resulting mass--metallicity relation at 1Gyr intervals resulting from the \textsc{ProSpect} fits. At each interval, the running median is indicated with a solid black line and black points, and the 1$\sigma$ range in the scatter is shown in dotted black lines. Where possible, we have included observational measurements at the relevant epochs as a comparison. These studies include those of \citet{Savaglio05}, \citet{Erb06}, \citet{Maiolino08}, \citet{Mannucci09}, \citet{Henry13a}, \citet{Henry13b}, \citet{LaraLopez13c}, \citet{Yabe14}, \citet{Zahid14}, \citet{Ly16}, \citet{Huang19}, \citet{Cameron19}, \citet{Horstman21}, \citet{Weldon20}, \citet{Sanders20}, \citet{Gillman21}. For measurements of the MZR that span a large redshift range (including those of \citealt{Ly16}), we plot the same values over multiple subpanels, and show the data using open symbols. }
	\label{fig:MassMetallicityEvolution}
\end{figure*}

\begin{table*}
	\centering
	\caption[Indicators]{Metallicity indicators and calibrations applied in each of the observational studies presented in Fig. \ref{fig:MassMetallicityEvolution}. PP04 refers to \citet{Pettini04}, KK04 refers to \citet{Kobulnicky04}, M08 refers to \citet{Maiolino08}. T$_e$ refers to the electron temperature, which is derived using auroral lines, and is regarded as a ``direct" method of measuring metallicity.  }
	\label{tab:Indicators2}
	\begin{tabular}{@{}c | cc | p{4cm} }
		\hline
		Study & Parameters/Emission lines & Calibration  & Comments \\
		\hline
		\citet{Savaglio05} & R$_{23}$, O$_{32}$ & KK04 &Strong line derivation\\
		\citet{Erb06} & N2 & PP04 &Strong line derivation\\	
		\citet{Maiolino08} & [\ion{O}{iii}]$\lambda$5007/H$\beta$, [\ion{O}{iii}]$\lambda$5007/[\ion{O}{ii}]$\lambda$3237 & \citet{Nagao06} & T$_e$ at low-$Z$, and photoionisation at high-$Z$ \\	
		\citet{Mannucci09} &R$_{23}$, [\ion{O}{iii}]$\lambda$5007/H$\beta$  &M08 &\\
		\citet{Yabe14} & N2 & PP04 &Strong line derivation\\
		\citet{LaraLopez13c} & O3N2 & PP04 &Strong line derivation\\
		\citet{Henry13a} & R$_{23}$ & KK04 &Strong line derivation\\
		\citet{Henry13b} & R$_{23}$ & M08 & Strong line derivation \\
		\citet{Zahid14} & N2 & PP04 &Strong line derivation\\
		\citet{Ly16} &[\ion{O}{ii}]$\lambda$3726,$\lambda$3729/H$\beta$, [\ion{O}{iii}]$\lambda$4959,$\lambda$5007/H$\beta$  &  & T$_e$ derivation\\
		\citet{Cameron19} & O3N2 & PP04 &Strong line derivation\\
		\citet{Huang19} & R$_{23}$, O$_{32}$ & KK04 & Strong line derivation\\
		\citet{Horstman21} & N2, O3N2 & PP04 & Strong line derivation\\
		\citet{Weldon20} & [\ion{O}{ii}]$\lambda$3726,$\lambda$3729/H$\beta$, [\ion{O}{iii}]$\lambda$4959,$\lambda$5007/H$\beta$ &  & T$_e$ derivation\\
		\citet{Sanders20} &[\ion{O}{iii}]$\lambda$5007/H$\beta$, O$_{32}$, [\ion{Ne}{iii}]$\lambda$3869/[\ion{O}{ii}]$\lambda$3727 & \citet{Bian18} & Strong line derivation \\
		\citet{Gillman21} & N2 & PP04 & Strong line derivation \\
		\hline		
	\end{tabular}
\end{table*}

The changing mass-metallicity relation at 1 Gyr intervals is shown in each subpanel of Fig. \ref{fig:MassMetallicityEvolution}. 
For each time interval, the median and 1$\sigma$ values are shown in solid and dashed black lines respectively. In generating the median and $1\sigma$ ranges, we demand that each bin includes at least 300 galaxies. 
For comparison, in each subpanel we show the $z\sim0$ MZR in cyan. 
Wherever possible, we have included observations of the mass-metallicity relation at the relevant epochs. 
These studies include \citet{Savaglio05}, \citet{Erb06}, \citet{Maiolino08}, \citet{Mannucci09}, \citet{Henry13a}, \citet{Henry13b}, \citet{LaraLopez13c}, \citet{Yabe14}, \citet{Zahid14}, \citet{Ly16}, \citet{Huang19}, \citet{Cameron19}, \citet{Weldon20}, \citet{Gillman21}, converting from the published $12+\log(\rm{O}/\rm{H})$ values to absolute metallicity via:
\begin{equation}
Z_{\rm{gas}}= Z_{\odot} \times 10^{[12+\log(\rm{O}/\rm{H})]-[12+\log(\rm{O}/\rm{H})]_{\odot}}, 
\end{equation} 
where $ Z_{\odot}=0.0142$\footnote{Note that this value differs from the commonly-assumed value of 0.02. } and $[12+\log(\rm{O}/\rm{H})]_{\odot} = 8.69$ \citep{Asplund09}. 
We additionally correct all stellar mass measurements to be consistent with a \citet{Chabrier03} IMF, employing correction factors as derived by \citet{Driver13}\footnote{To convert from a \citet{Salpeter55} IMF we multiply by a factor of 0.65, and for a \citet{Baldry03} IMF we use a factor of 1.2. }. 

A number of different methods were used to make the observed metallicity measurements presented in Fig. \ref{fig:MassMetallicityEvolution}. 
The R$_{23}$ parameter \citep{Pagel79} and the N2 parameter were used, as well as the O3N2 parameter and the O$_{32}$ parameter. 
Even if the same parameters are applied to measure metallicity, they can be differently calibrated. 
The definitions of these parameters are: 
\begin{align}
\rm{R}_{23} &\equiv \frac{[\rm{O\,\scriptstyle II}]\lambda 3727 + [\rm{O\,\scriptstyle III}]\lambda 4959,\lambda 5007}{\rm{H}\beta} \\
\rm{O}_{32} &\equiv \frac{[\rm{O\,\scriptstyle III}]\lambda 4959 + [\rm{O\,\scriptstyle III}]\lambda 5007}{[\rm{O\,\scriptstyle II}]\lambda 3727} \\
\rm{N2} &\equiv \log\left(\frac{[\rm{N\,\scriptstyle II}]\lambda 6584}{\rm{H}\alpha}\right) \\
\rm{O3N2} &\equiv \log\left(\frac{[\rm{O\,\scriptstyle III}]\lambda5007/\rm{H}\beta}{[\rm{N\,\scriptstyle II}]\lambda  6584/\rm{H}\alpha}\right) 
\end{align}

Frequently applied calibrations include the N2 and O3N2 calibrations by \citet{Pettini04}, which were later updated by \citet{Tremonti04}, and the calibration of the R$_{23}$ parameter by \citet{Kobulnicky04}.
We summarise the parameters and calibrators used by each of these studies in Table \ref{tab:Indicators2}. Note that in the high-mass regime, strong line derivations are generally employed, whereas in the low-mas regime, electron temperature-based derivations are employed. 
Some scatter between the observations themselves is to be expected due to these different indicators alone, as highlighted by \citet{Kewley08, LopezSanchez10b, LopezSanchez12}. 

The agreement between the trends recovered by observations and our forensically-determined MZR at each epoch is remarkable, at all stellar mass ranges. 
Note that in classical SED fitting approaches, where metallicity is assumed to be constant with time, the inferred metallicity distribution would be recovered to be constant across all subpanels of Fig. \ref{fig:MassMetallicityEvolution}, with only stellar masses evolving with time. 
This would be in clear tension with observations, highlighting the improvement that has been gained through the implementation of an evolving metallicity parametrisation in our modelling. 
We note that the metallicities derived for the most massive galaxies in our sample ($M_* > 10^{10.5} \rm{M}_{\odot}$) tend to be slightly higher than observed metallicities at these stellar masses, and hence the flattening of our derived MZR is slightly weaker than observed relations. 
Observations between 9-11 Gyr \citep{Yabe14, Zahid14, Erb06, Gillman21} recover metallicities lower than our forensic values in the highest stellar mass bins.
This looks likely to be the result of metallicity saturation from the N2 indicator, which is known to occur for metallicities above $12+\log(\rm{O}/\rm{H}) \geq 8.8$ (corresponding to $Z_{\rm gas}\geq 0.018$), as discussed by \citet{LaraLopez13c}. 
Similarly, in the 7 Gyr bin, the observations of both \citet{Ly16} and \citet{Weldon20} flatten at higher stellar masses, whereas other metallicity measurements focussing on a larger mass range do recover significantly higher metallicities in the overlapping mass range. This also suggests that saturation is occurring when using T$_e$-based metallicity measurements. 

The potential presence of saturation in observational metallicity measurements makes a comparison between our values and observations for massive galaxies at 9-11 Gyr potentially biased.

\subsubsection{Evolving shape and normalisation}

\begin{figure}
	\centering
	\includegraphics[width=85mm]{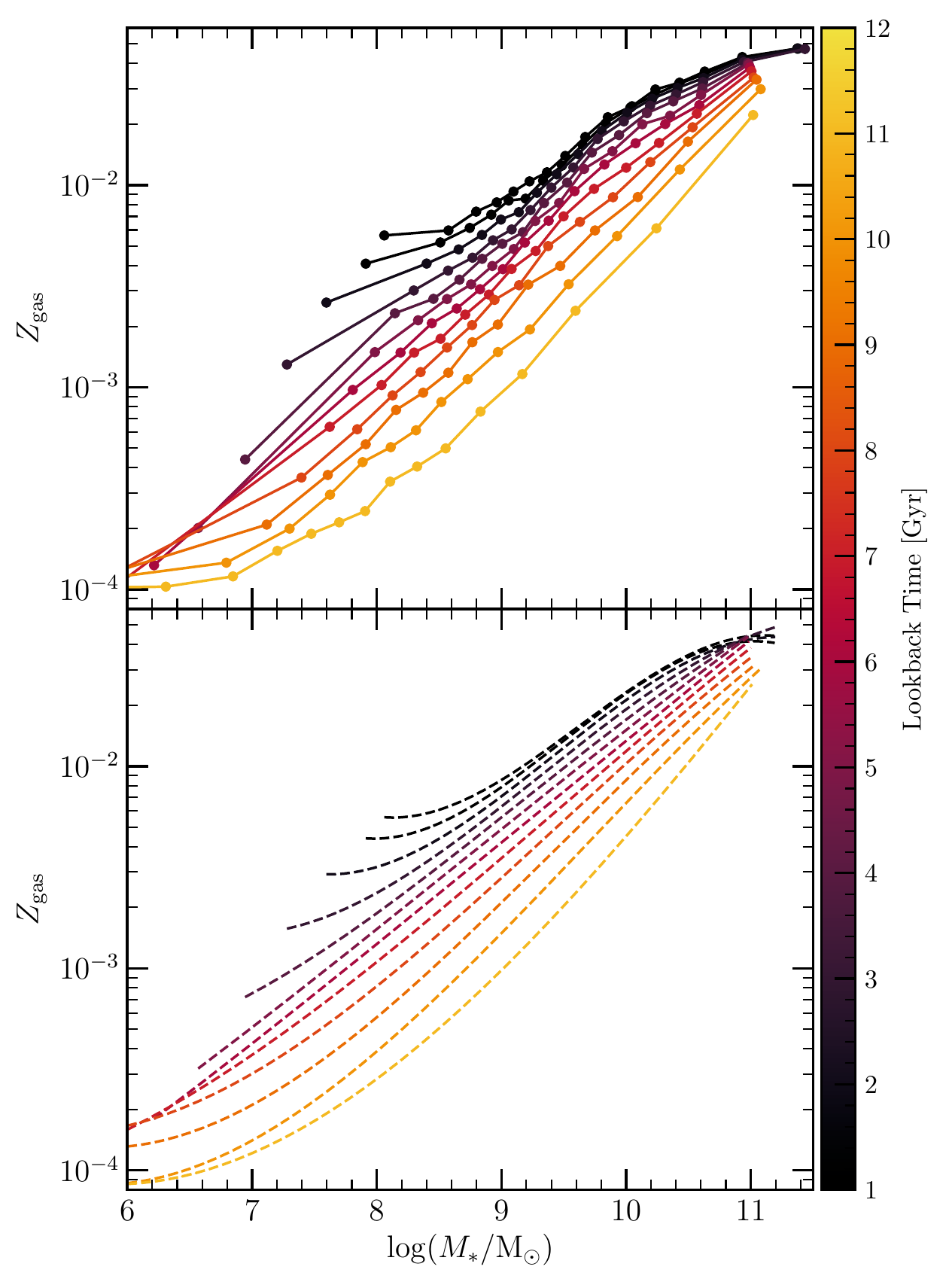}
	\caption{The top panel shows the median MZR measured at each epoch (as shown in each subpanel of Fig. \ref{fig:MassMetallicityEvolution}), coloured by lookback time. The bottom panel shows the functional form of the evoving MZR, as given by Equation \ref{eqn:MZRmedian}.  }
	\label{fig:MassMetallicityEvolution_median}
\end{figure}

To visually demonstrate how the form of the MZR evolves with time, we present the median MZR at each epoch in Fig. \ref{fig:MassMetallicityEvolution_median}, coloured by lookback time. As expected, the MZR reduces in normalisation with increasing lookback time, with the bending at higher masses becoming less pronounced at earlier times. This changing shape (albeit subtle) seems to suggest a ``saturation" of metallicities occurring in galaxies at recent times. 

This evolution in the metallicity can be described as a function of stellar mass and lookback time ($t_{\rm lb}$) by:

\begin{equation}
	\log(Z_{\rm gas})(M_*, t_{\rm lb}) = \sum_{i=0}^{3} f_i(t_{\rm lb})m^i
\label{eqn:MZRmedian}
\end{equation}
where
\begin{align}
	m &= \log(M_*/\rm{M}_{\odot}) - 10\\	
	f_i(t_{\rm lb}) &= \sum_{j=0}^{5} a_{i,j}t_{\rm lb}^j,
\end{align}
and the $a_{i,j}$ coefficients are provided in Table \ref{tab:Indices}. These fits are shown in the lower panel of Fig. \ref{fig:MassMetallicityEvolution_median}. 
The scatter (in dex) can be described by the same functional form. The corresponding coefficients are also shown in  Table \ref{tab:Indices}. 

\begin{table*}
	\centering
	\caption[Indices]{Coefficients to describe the time evolution of the MZR median and scatter, as used in Equation \ref{eqn:MZRmedian}. }
	\label{tab:Indices}
	\begin{tabular}{@{}l | rrrrrr}
\hline		
\multicolumn{7}{c}{\text{Median evolution coefficients}} \\
		& $a_{i,0}$  & $a_{i,1}$   & $a_{i,2}$    & $a_{i,3}$  & $a_{i,4}$    & $a_{i,5}$  \\
		\hline
$a_{0,j}$ & $-1.67$              & $ 6.32\times10^{-2}$ & $-3.08\times10^{-2}$ & $ 3.62\times10^{-3}$ & $-1.57\times10^{-4}$ &  0         \\
$a_{1,j}$ & $ 4.25\times10^{-1}$ & $-3.69\times10^{-3}$ & $ 3.86\times10^{-3}$ & $-4.16\times10^{-4}$ & $ 2.38\times10^{-5}$ &  0         \\
$a_{2,j}$ & $ 1.42\times10^{-2}$ & $-1.38\times10^{-1}$ & $ 5.24\times10^{-2}$ & $-6.88\times10^{-3}$ & $ 3.37\times10^{-4}$ & $-3.78\times10^{-6}$ \\
$a_{3,j}$ & $-2.67\times10^{-2}$ & $-7.20\times10^{-2}$ & $ 3.58\times10^{-2}$ & $-6.00\times10^{-3}$ & $ 4.20\times10^{-4}$ & $-1.05\times10^{-5}$ \\
\hline	
\hline	
\multicolumn{7}{c}{\text{Scatter evolution coefficients}} \\
		& $a_{i,0}$  & $a_{i,1}$   & $a_{i,2}$    & $a_{i,3}$  & $a_{i,4}$    & $a_{i,5}$  \\
		\hline
$a_{0,j}$ & $-3.65\times10^{-1}$ & $-5.85\times10^{-3}$ & $ 3.56\times10^{-4}$ & $1.04\times10^{-4} $ & $ 1.61\times10^{-7}$ &   0       \\
$a_{1,j}$ & $-3.34\times10^{-1}$ & $ 4.30\times10^{-2}$ & $-8.47\times10^{-3}$ & $9.06\times10^{-4} $ & $-2.92\times10^{-5}$ &   0       \\
$a_{2,j}$ & $-1.66\times10^{-1}$ & $ 4.26\times10^{-2}$ & $-3.29\times10^{-2}$ & $8.93\times10^{-3} $ & $-8.92\times10^{-4}$ & $2.99\times10^{-5}$ \\
$a_{3,j}$ & $-1.05\times10^{-2}$ & $ 5.16\times10^{-3}$ & $-1.62\times10^{-2}$ & $4.92\times10^{-3} $ & $-5.05\times10^{-4}$ & $1.72\times10^{-5}$ \\
		\hline		
	\end{tabular}
\end{table*}

The evolution of the MZR shown in Fig. \ref{fig:MassMetallicityEvolution_median} shows that the chemical enrichment of a $10^{8}\,\rm{M}_{\odot}$ galaxy is significantly more different at high-$z$ versus low-$z$, than that of a $10^{10}\,\rm{M}_{\odot}$ galaxy, which displays less variation in metallicity with cosmic time. 
This behaviour is attributed to the ``galaxy downsizing" phenomenon that is recovered in this analysis (as highlighted by \citealt{Bellstedt20b}). 
Massive galaxies tend to assemble earlier, and hence their late-time chemical enrichment is minimal, whereas low-mass galaxies continue to form stellar mass at the present day. Consequently, this stellar mass range experiences more prolonged chemical enrichment.

\subsection{Trends with SFR}

\begin{figure*}
	\centering
	\includegraphics[width=180mm]{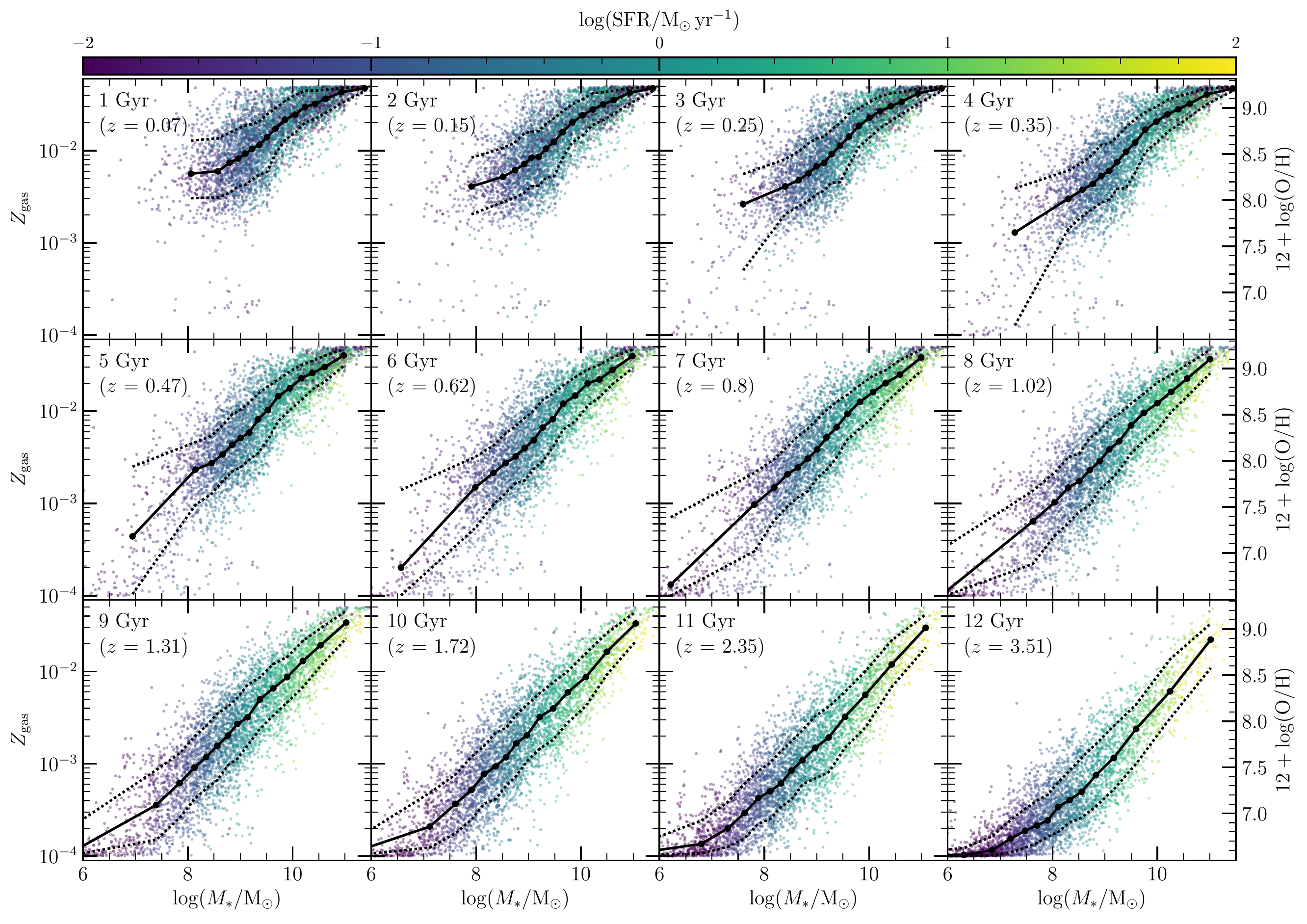}
	\caption{The mass--metallicity relation in 1 Gyr intervals for the same galaxy population, coloured by the SFR at that epoch. As in Fig. \ref{fig:MassMetallicityEvolution}, the median and $1\sigma$ MZR are shown in solid and dashed black lines in each subpanel. }
	\label{fig:MassMetallicityEvolution_SFR}
\end{figure*}

For each cosmic epoch shown in Fig. \ref{fig:MassMetallicityEvolution}, we now assess how the galaxy SFRs change across the mass-metallicity relation. 
In each subpanel in Fig. \ref{fig:MassMetallicityEvolution_SFR}, each galaxy is now coloured by $\log(\rm SFR)$, with galaxies lighter in colour representing objects that are more rapidly forming stars. 
Fig. \ref{fig:MassMetallicityEvolution_SFR} shows clearly that the position of a galaxy within the MZR is highly dependent on the star formation rate. 
In particular, we highlight that this trend is strongest at earlier times, and becomes less distinct in the last 5 Gyr, where galaxies start to saturate in gas-phase metallicity. 
This plot shows that in the early Universe, metallicity enrichment and SFR are directly linked to stellar mass, such that at fixed stellar mass, the SFR is relatively constant for different metallicity values. 
A vertical gradient in colour (a dependence of SFR on metallicity), only starts becoming apparent at a lookback time of 10 Gyr --- the same epoch in which the bending of the MZR starts to become apparent. 

It is conceivable that the recovered behaviour is simply a consequence of the parametrisation of SFHs that we implement, and that therefore the trends in the early Universe are highly simplified. Additionally, the actual constraint on the SFH and ZH at such large lookback times is also very small when applying SED fitting, so there may not actually be any real signal here - only what our models are telling us.
Note however, that if this behaviour really is simply the consequence of the adopted SFH, then the fact that the evolving MZR is so well recovered provides confidence that the adopted SFH and ZH parametrisation is indeed appropriate. 

We highlight that Fig. \ref{fig:MassMetallicityEvolution_SFR} also conveys the clear mass-dependence of the CSFH. In particular, at all redshifts, the highest-mass systems have the greatest SFRs. The value of this maximum SFR changes with redshift, in accordance with the overall decline in the CSFH over the past $\sim$ 10 Gyr \citep[as shown by][]{Madau14, Bellstedt20b}. This is evident when observing the changing SFR within galaxies at a fixed stellar mass over time: At a lookback time of 11 Gyr, galaxies with $M_*=10^{10}\rm{M}_{\odot}$ have SFRs $\sim10\,\rm{M}_{\odot}\rm{yr}^{-1}$, whereas by 1 Gyr lookback time the star-forming galaxies in this mass range have SFRs $\sim1\,\rm{M}_{\odot}\rm{yr}^{-1}$. 

We do not apply a selection in Fig. \ref{fig:MassMetallicityEvolution_SFR} based on sSFR, which means the plot features both star-forming, and quenched galaxies. The presence of quenched/quenching systems is first apparent at a lookback time of 10 Gyr, where the first high-mass galaxies --- $\log(M_*/\rm{M}_{\odot}) > 10$ --- are becoming dark purple in colour, corresponding to $\log(\rm{SFR}/\rm{M}_{\odot}\rm{yr}^{-1}) < -1$. By 7-8 Gyr lookback time, these objects become much more prevalent. The presence of these system causes a noticeable visual dilution of the SFR trends, which is maximised by 1 Gyr, when a significant fraction of the high-mass galaxies is quenched. As such, there is only a minimal trend with SFR in the 1 Gyr panel of this plot. 

\begin{figure*}
	\centering
	\includegraphics[width=180mm]{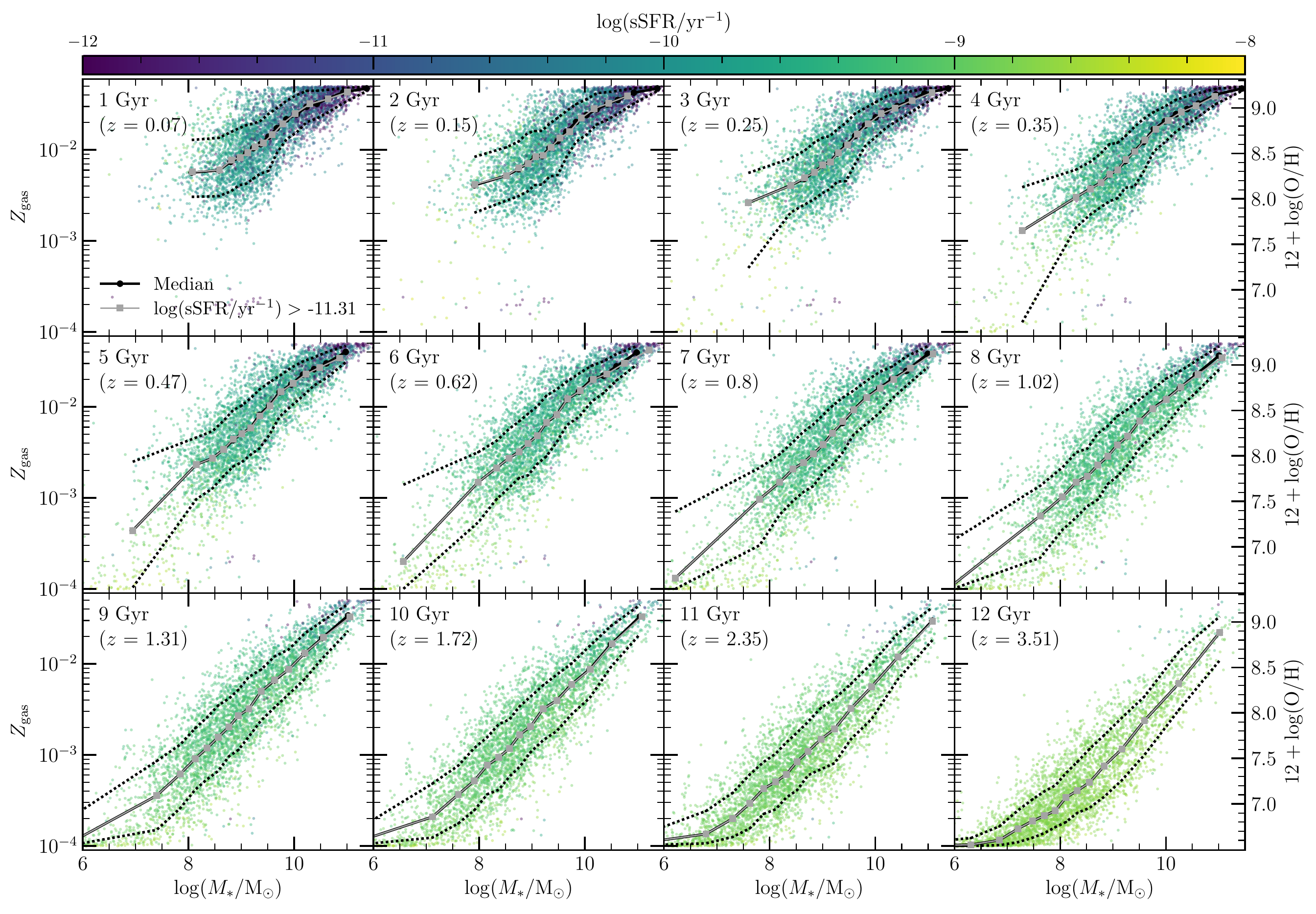}
	\caption{The mass--metallicity relation in 1 Gyr intervals for the same galaxy population, coloured by the sSFR at that epoch. As in Fig. \ref{fig:MassMetallicityEvolution}, the median and $1\sigma$ MZR are shown in solid and dashed black lines in each subpanel. The running median for the star-forming population (as defined by $\log(\rm sSFR/\rm{yr}^{-1})>-11.31$) is shown in grey.   }
	\label{fig:MassMetallicityEvolution_sSFR}
\end{figure*}

To assess how the trends in Fig. \ref{fig:MassMetallicityEvolution_SFR}  change with sSFR, we present the same parameter space coloured by sSFR in Fig. \ref{fig:MassMetallicityEvolution_sSFR}. 
In addition to the running median in each subpanel of the total population, we include the running medianfor the star-forming population (as defined by $\log(\rm sSFR/\rm{yr}^{-1})>-11.31$, \citealt{Davies19}). 
We highlight here that the consequence of including passive systems in our analysis are essentially negligible.

An important caveat to note, that will impact the trends shown in Fig. \ref{fig:MassMetallicityEvolution_SFR}, is that we neglect the impact of potential historical mergers in this analysis. If a galaxy has undergone a major merger in the past, then instead of a single progenitor galaxy, there would be two. As a result, for this object we are likely to have overestimated the mass (and, by extension, the SFR) of the galaxy at high lookback times. It is therefore possible that the high-mass high-SFR objects in the panels at high lookback times could be an artefact of this assumption, and that in reality such galaxies do not exist. 

\section{The Mass--Metallicity--SFR plane}
\label{sec:Plane}

\begin{figure}
	\centering
	\animategraphics[width=\columnwidth, autoplay, loop, controls]{6}{figs/GIF/}{0}{60}
	\caption{The mass--metallicity--SFR plane as determined by the fitted parameters. The values in 3D space, as well as their projections onto the three 2D spaces, are all plotted. Colours indicate the orthogonal distance of the point to the place. 3D values are shown in red/orange hues, while their corresponding projections are shown in blue/green hues, to allow them to be distinguished. The fit made to the plane by \citet{LaraLopez13a} is shown in orange. The tilt of the plane in this fit is significantly larger than the tilt we measure. Note that the disagreement is greatest at low metallicities, where observations are sparse. The cyan surface is the measurement by \citet{Mannucci10}, and the blue surface is the fit to the 3D space by \citet{Curti20}. Here, the surface has a significantly larger tilt, and the saturation metallicity is lower than the maximum metallicities we recover. See online version for 3D video.  }
	\label{fig:Plane}
\end{figure}

It is now understood that the mass--metallicity relation is a projection of the 3D mass--metallicity--SFR plane \citep{Magrini12, LaraLopez13a, Peeples13}. Using the outputs of SED fitting, we can present our data in this plane, with a broad range of stellar masses, metallicities and SFRs. 

To fit the plane in three dimensions, we employ the \textsc{R} package \textsc{Hyperfit} \citep{Hyperfit}. When fitting a plane to our derived values of stellar mass, SFR and metallicity, we have converted our absolute metallicity values to oxygen abundances, for the sake of easy comparison against observations. We additionally restrict our fit to galaxies with specific SFR values $\log(\rm{sSFR}/\rm{yr}^{-1}) > -11.5$, so that we do not include quenched and quenching galaxies in our fit.

Our fit to the $z\sim0$ Mass--Metallicity--SFR plane can be described with the functional form: 
\begin{equation}
[12+\log(\rm{O}/\rm{H})] = \alpha \left(\frac {\rm{SFR} }   {\rm{M}_{\odot}\rm{yr}^{-1}}     \right)  + \beta  \log\left(\frac{M_*}{\rm{M}_{\odot}}\right) + \gamma
\end{equation}
where
\begin{align*}
\alpha&=-1.426\\
\beta&=1.628\\
\gamma&=-7.592
\end{align*}

In Fig. \ref{fig:Plane}, we plot the plane as fitted by \citet{LaraLopez13a} in orange. This plane intersects with ours at high masses and metallicities, however it diverges in the lower-mass  and lower-metallicity regimes. This is a consequence of the \citet{LaraLopez13a} sample, which focussed on galaxies with stellar masses of $\sim10^{9}\rm{M}_{\odot}$, and as such a discrepancy of the plane at low stellar masses is unsurprising. 
We additionally present the 3D fitted surface by \citet{Mannucci10} in cyan, which only covers the high-mass, high-metallicity regime of the 3D space. Similarly, we show the 3D surface fitted by \citet{Curti20} is shown in blue. 
We find that the surfaces from \citet{Mannucci10} and \citet{Curti20} have a signficantly stronger tilt than ours, and also saturate at a much lower metallicity that the maximum metallicities we derive. 
The structure in the surfaces fitted by both \citet{Mannucci10} and \citet{Curti20} do not visually provide a better fit to the mass--metallicity--SFR relation we derive than a simple plane does. 
In fact, the plane derived by \citet{LaraLopez13a} is in much better agreement with the distribution of our derived data than either of the two other surfaces. 

We highlight here that, because the \texttt{Zfinal} parameter is a free parameter in our implementation, the planar structure of our mass--metallicity--SFR plane was not prescribed by our evolving metallicity model. 

\subsection{Evolving plane}

Our forensic analysis also allows us to measure the evolution of the Mass--Metallicity--SFR plane. The evolution of the planes with cosmic time is shown in Fig. \ref{fig:PlaneEvolving}. Each subpanel shows the mass--metallicity--SFR plane in a specific lookback time interval. The projection of the points onto each of the three related two-dimensional spaces (i.e mass--metallicity, mass--SFR and SFR--metallicity) is indicated on the edges of the plot with blue/green points. A consequence of the functional form of the star formation history, is that the early build-up of galaxies in this sample is very similar, due to the truncation of each SFH at early times. As a result, the main sequence (the 2D projection at the bottom of each subpanel) can be seen to be extremely narrow at high lookback times. At this epoch, the resulting stellar masses are limited by the SFR, and hence they are very tightly correlated.

\begin{figure*}
	\centering
	\animategraphics[width=180mm, autoplay, loop, controls]{6}{figs/GIF2/}{0}{23}
	\caption{The Mass--Metallicity--SFR plane at regular lookback time intervals between 1 and 11 Gyr. The values themselves are plotted in red/orange hues, while their corresponding projections onto the three 2D spaces is plotted in blue/green hues, for clarity. As in Fig. \ref{fig:Plane}, points are coloured according to their orthogonal distance from the fitted plane. The plane at each epoch is shown in solid grey, and for comparison the $z\sim0$ plane is indicated with the dashed line for reference. The tilting of the mass--metallicity--SFR plane with cosmic time is evident. Additionally, it is clear that the scatter around the plane increases with cosmic time. See online version for 3D video.  }
	\label{fig:PlaneEvolving}
\end{figure*}

The plane can be seen to evolve with cosmic time. In particular, with time the plane tilts toward high-mass, low-SFR systems with increasing time. The low-metallicity portion of the plane tilts toward higher SFRs, however this is unconstrained at recent times due to the lower mass limit of our observations. This evolution is most notable in lookback times beyond 7 Gyr, in the 9 and 11 Gyr panels. We highlight that, as mentioned above, this epoch is highly influenced by our choice of SFH parametrisation, and hence the measurement of this evolution must be assessed with this caveat. 

The evolution of the plane with lookback time can be expressed as:
\begin{align}
[12+\log(\rm{O}/\rm{H})] &= \alpha(t_{\rm{lb}}) \log \left(\frac {\rm{SFR} }   {\rm{M}_{\odot}\rm{yr}^{-1}}     \right) \nonumber \\
&  + \beta(t_{\rm{lb}})  \log\left(\frac{M_*}{\rm{M}_{\odot}}\right) + \gamma(t_{\rm{lb}})  
\label{eqn:Plane}
\end{align}
where
\begin{align*}
\alpha(t_{\rm{lb}})&=-0.9202  + 0.2916 t_{\rm{lb}} -0.05082 t_{\rm{lb}}^2\\
\beta(t_{\rm{lb}})&=1.198   -0.2658t _{\rm{lb}} +   0.04884 t_{\rm{lb}}^2\\
\gamma(t_{\rm{lb}})&=-2.785 +  2.268 t_{\rm{lb}} -0.4267 t_{\rm{lb}}^2
\end{align*}
Here, $t_{\rm{lb}}$ is the lookback time in Gyr. The scatter in the plane is seen to evolve as:
\begin{align*}
\sigma(t_{\rm{lb}})&=0.427  -0.02979  t_{\rm{lb}} + 0.0026182 t_{\rm{lb}}^2
\end{align*}
This fit to the evolving plane parameters is shown in Fig. \ref{fig:EvolvingPlaneParameters}. 

In an analysis of stellar masses, SFRs and metallicities spanning a wide redshift range, \citet{Mannucci10} concluded that there was no evolution in the mass-metallicity--SFR plane from $z=0$ to $z=2.5$. This is in constrast with the subtle evolution that we derive in our analysis. 
Note that the study by \citet{Mannucci10} focussed on galaxies with $M_*>10^{9.5}\rm{M}_{\odot}$, and hence the stellar mass range may well have been too small to detect the evolution that we infer. 
\citet{LaraLopez10b} also concluded in their analysis that there was no evidence for an evolution in the plane.

\section{Discussion}
\label{sec:Discussion}

There are numerous caveats associated with the analysis presented in this work, both in the manner in which our values are compared against the observed metallicity values, and also inherent to the modelling method we have applied. 
We discuss these issues in this section. 

A major challenge associated with any comparison of modelled metallicities to spectroscopically-derived values, is that there are significant biases that accompany spectroscopic measurements. 
This is the result of relying on different emission lines between different datasets, different strong line parameters, and finally different calibration methods to transform the strong line measurements into abundances. 
This challenge was highlighted by \citet{Sanchez19b}, who circumvented some of these systematics by electing to use 11 different calibrators in their analysis, rather than only selecting one. 
The resulting mass--metallicity relations that they derive not only vary in normalisation by up to 0.5 dex, but the resulting slopes also vary dramatically. 
A similar depiction of this challenge was presented earlier by \citet{Kewley08, LopezSanchez10b, LopezSanchez12}. 

Not only is measuring the absolute oxygen abundance challenging, but there is extra uncertainty associated with the conversion of oxygen abundance to total metallicity, which is the quantity we model in our analysis.  
This was highlighted by \citet{Gallazzi05} who, using stellar population models to infer the total metallicities of SDSS spectra, compared the total metallicities against oxygen abundances determined for the same sample as \citet{Tremonti04}. 
While there was a clear correlation between these parameters, the scatter was also significant. 
Consequently, scatter between the observed and our modelled metallicities could also arise from the scaling between oxygen abundances and total metallicities.

A key simplification in the SED fitting approach that we have taken, is that mergers that may have occurred in a galaxy's past are ignored.
It would be reasonable to assume that a massive galaxy that only has a single progenitor that has epochs of extremely high star formation would likely have different metallicities to a massive galaxy that formed as the result of a series of major mergers, each progenitor of which may have had smaller star formation rates. 
In our approach, we would not differentiate such systems, as we simply model the history of all the stars currently present in a galaxy. In an analysis focussing on the metallicities of merging galaxies, \citet{Horstman21} found that galaxies in mergers had suppressed metallicities versus isolated galaxies at the same stellar mass at $2<z<2.7$, highlighting that mergers will definitely have an impact on the metallicity of an object.
Additionally, the associated stellar masses and star formation rates are likely to be biased to be high in our analysis. The extent of this bias would depend on the merger rates throughout cosmic time. A quantification of this potential bias is beyond the scope of this work, however this is a likely explanation for the very high-mass objects presented in our work at earlier lookback times, that are higher mass than the typical galaxy observations at that epoch. 

Furthermore, our SED fitting approach does not explicitly account for any interactions between a galaxy and its surroundings, including phenomena such as feedback (either stellar or AGN), or any gas inflows. These are mechanisms that could have the effect of either enriching or diluting the metal content of gas in galaxies \citep[as demonstrated by, for example, ][]{LopezSanchez15}. Some allowance for the occurrence of gas inflows is provided by the linear metallicity evolution implementation, unlike the closed-box model (which was utilised in the main body of \citealt{Bellstedt20b}, see \citealt{Robotham20} for further details), however this is not explicit.
Work by \citet{Rupke10} demonstrated that the metallicity gradients of galaxies undergoing interactions are significantly flatter than non-interacting galaxies caused by interaction-induced gas inflow. Note that in our approach we cannot model the spatially-resolved nature of metallicity within galaxies. 
\citet{Chisholm18} looked at outflows as a mechanism for producing the shape of the MZR. 
Based on an extensive modelling of UV spectral observations for 7 local galaxies, \citet{Chisholm18} determined that the metal outflow rate in galaxies linearly correlates with their stellar mass. 
Outflows were also invoked in the theoretical analysis of \citet{Dayal13} to describe the fundamental metallicity relation. 
Contrastingly, \citet{Calura09} determined that winds were not required to explain the shape of the MZR, and that it could instead be explained by the lower star formation efficiency of lower-mass systems. 
The analysis presented in this work suggests that an explicit treatment of outflows is not required to reproduce the general shape of the MZR. 

An additional phenomenon that is unconsidered in our approach is that of galaxy starvation, in which the gas inflow to a galaxy is disrupted, leading to a gradual truncation of star formation within a galaxy. 
This process was discussed in detail by \citet{Peng15}, who in particular noted that in a starvation scenario, the final star formation epoch is likely to be more chemically enriched than any previous star formation, due to the lack of dilution caused by gas inflows. 
This effect is predicted to be responsible for the discrepancy between \textit{stellar} metallicities of passive and star-forming galaxies at fixed stellar mass (as presented by, for example, \citealt{Peng15, Trussler20}). 
Such a late-time acceleration of chemical enrichment of galaxies would not be considered in our approach.

While the present-day gas-phase metallicities of our galaxies are free parameters within our analysis, the metallicity at the beginning of each galaxy's history is fixed to the lower limit of the \citet{Bruzual03} stellar population templates. It is conceivable that a galaxy forming later in the Universe is formed out of non-pristine gas, and therefore we might be underestimating its initial metallicity. Contrastingly, this lower metallicity limit of $10^{-4}$ is likely too high for galaxies that form very early in the Universe, where the formation gas is likely to be pristine. It is unknown to what extent these simplifications impact the metallicity evolution of galaxies that we infer, however we point the reader to the discussion in Appendix \ref{sec:LinearityValidity}, where we highlight a potential consequence of the initial metallicity value on the assumption of linearity in the metallicity evolution prescription.

\section{Conclusions}

SED fitting techniques are widespread in use, and are frequently employed to measure the stellar masses of galaxies. Historically, SED fitting techniques have employed a simple parametrisation of the star formation history, and have typically modelled the corresponding metallicity history with a single value held constant over the history of the galaxy. This simplification has a significant impact on not only the star formation histories of galaxies \citep[as discussed in][]{Bellstedt20b}, but as a consequence the derived stellar masses can also be affected. Using the SED fitting code \textsc{ProSpect} \citep{Robotham20}, we have used a simple prescription to model an evolving metallicity for individual galaxies \citep[used in][to extract a cosmic star formation history consistent with observational measurements]{Bellstedt20b} to derive the mass--metallicity relation for a sample of $\sim$4,500 $z<0.06$ galaxies form the GAMA survey. 

Despite the caveats discussed in the previous section, the agreement that we observe between our inferred mass--metallicity relation and observations across a wide range of stellar masses, metallicities and cosmic time is remarkable.  We show the evolution of the median MZR with cosmic time, and show this evolution in a functional form in Equation \ref{eqn:MZRmedian}. 

Furthermore, we use consistently derived metallicities, stellar masses and SFRs to make a planar fit to the 3D mass--metallicity--SFR space. We show that this space can be well described by a plane over cosmic time, and show that as galaxies increase in stellar mass and metallicity, and reduce the rate at which they are formig stars, the mass--metallicity--SFR plane tilts. This evolution mostly occurs at lookback times greater than 7 Gyrs. 
We present the evolution of this plane in functional form in Equation \ref{eqn:Plane}. 

Combined with the accurate cosmic star formation history derived using this implementation of \textsc{ProSpect} \citep[as presented in][]{Bellstedt20b}, the analysis presented in this work reaffirms that galaxy stellar populations can be modelled accurately using SED fitting of broadband photometry alone, if careful consideration is given to the evolution of the gas phase metallicity in addition to the evolution of the star formation history.

\section{Data Availability}

The \textsc{ProSpect} catalogue will be collated into a DMU, and can be accessed via a GAMA collaboration request\footnote{\url{http://www.gama-survey.org/collaborate/}}. 


\section{Acknowledgements}

SB thanks Mat\'{i}as Bravo for providing \textsc{Shark} data, and Dr Adam Stevens for many useful discussions. 

GAMA is a joint European-Australasian project based around a spectroscopic campaign using the Anglo-Australian Telescope. The GAMA input catalogue is based on data taken from the Sloan Digital Sky Survey and the UKIRT Infrared Deep Sky Survey. Complementary imaging of the GAMA regions is being obtained by a number of independent survey programmes including GALEX MIS, VST KiDS, VISTA VIKING, WISE, Herschel-ATLAS, GMRT and ASKAP providing UV to radio coverage. GAMA is funded by the STFC (UK), the ARC (Australia), the AAO, and the participating institutions. The GAMA website is \url{http://www.gama-survey.org/}.

SB and SPD acknowledge support by the Australian Research Council's funding scheme DP180103740. 
ASGR acknowledges support from the ARC Future Fellowship scheme (FT200100375), and LPD acknowledges support from the ARC Future Fellowship scheme (FT200100055).
This work was supported by resources provided by the \textit{Pawsey Supercomputing Centre} with funding from the Australian Government and the Government of Western Australia. We gratefully acknowledge \textit{DUG Technology} for their support and HPC services.

We have used \textsc{R} \citep{R} and \textsc{python} for our data analysis, and acknowledge the use of \textsc{Matplotlib} \citep{Hunter07} for the generation of plots in this paper. This research made use of \textsc{Astropy},\footnote{\url{http://www.astropy.org}} a community-developed core \textsc{python} package for astronomy \citep{Astropy13, Astropy18}, \textsc{Pandas} \citep{Pandas}, and \textsc{NumPy} \citep{NumPy}.

\bibliographystyle{mnras}
\setlength{\bibsep}{0.0pt}
\bibliography{BibLibrary}

\appendix

\section{Validity of proportionally evolving metallicities}
\label{sec:LinearityValidity}

\begin{figure*}
	\centering
	\includegraphics[width=180mm]{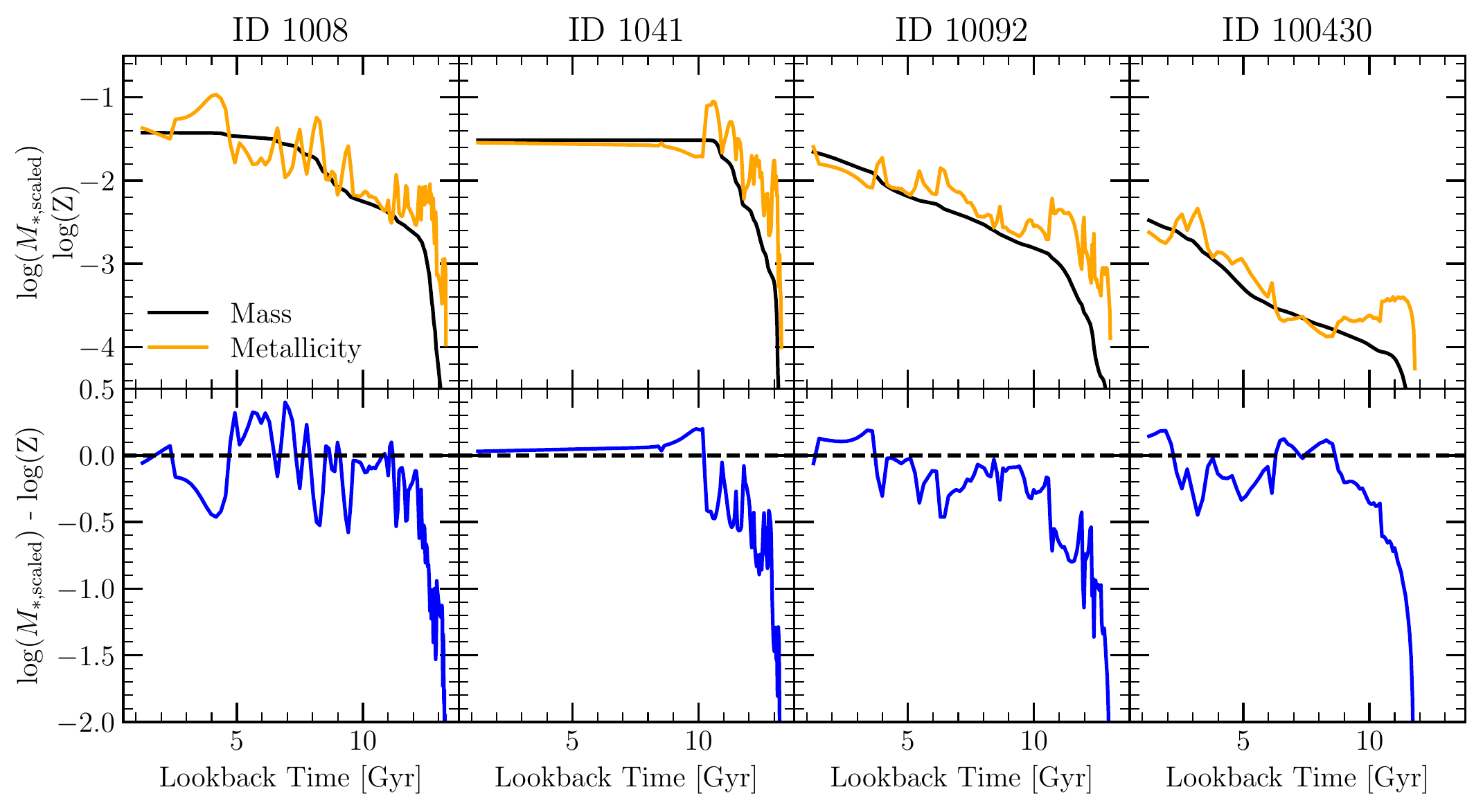}
	\caption{The relative stellar mass and metallicity build-up of four example \textsc{Shark} galaxies (top panel), and the resulting ratio between these (bottom). The dashed black line indicates proportional metallicity evolution.}
	\label{fig:SharkExamples}
\end{figure*}

In order to assess the validity of our ``linearly-evolving" metallicity evolution parametrisation, we turn to outputs from the semi-analytic model \textsc{Shark} \citep{Lagos18}. 
For a sample of $\sim$6000 galaxies sampled from one subvolume of the $z=0.07$ snapshot (Bravo et al. in prep), we extract the mass build-up, and gas-phase metallicity histories for each galaxy. 
In order to test how well these galaxies are, on average, approximated by our model, we need to identify the extent to which the mass build-up and metallicity evolution are proportional to each other. 
To do this we scale the mass profiles to the metallicity profiles until the differences between them are minimised, and then we quantify the difference between these profiles in log space. 
If a galaxy has purely proportional evolution, then this value would be 0 at all epochs. Some examples of how this behaves for galaxies in \textsc{Shark} are shown in Fig. \ref{fig:SharkExamples}. 
Note that the metallicity build-up does in general follow the stellar mass build-up fairly closely, with the greatest discrepancies typically displayed in the first few billion years of the galaxy's lifetime. 
Note that this discrepancy is apparent in the scaled Mass/Z ratio shown in the bottom panels of the figure.

Although the lower limit of gas-phase metallicity prescribed within \textsc{Shark} is $10^{-7}$, the value tends to be $\sim 10^{-4}$ as soon as star formation has started. 
This is close to the initial value of $10^{-4}$ that we are forced to implement due to the limit of the \citet{Bruzual03} stellar population templates.  

\begin{figure}
	\centering
	\includegraphics[width=85mm]{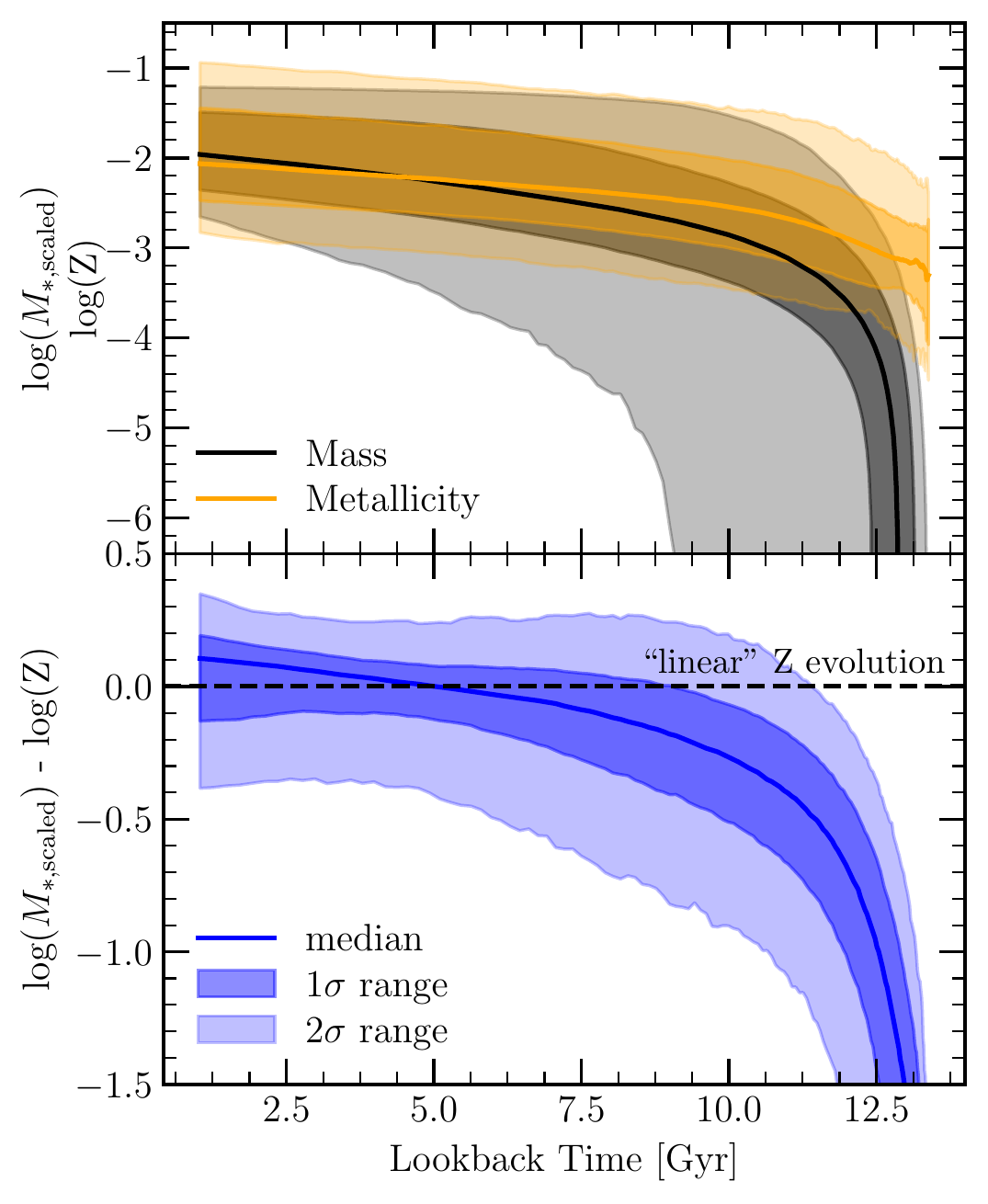}
	\caption{Top: Overall relative build-up of mass (black) and metallicity (orange) for a sample of $\sim$6000 galaxies from the semi-analytic model \textsc{Shark}. 
			Bottom: Distribution of the corresponding stellar mass build-up to gas-phase metallicity evolution ratios (blue, with shaded regions indicating the 1- and 2-$\sigma$ ranges). The dashed black line indicates the assumption of our proportional metallicity evolution (given by the \texttt{Zfunc\_massmap\_lin} function). Note that \textsc{Shark} is consistent with linear evolution to lookback times of $\sim$8 Gyr.   }
	\label{fig:LinearValidity}
\end{figure}

The result of this analysis over the full sample of galaxies is shown in Fig. \ref{fig:LinearValidity}. 
The relative mass and metallicity build-up are shown in the top panel, and their ratio is shown in  in the bottom panel, with the 1- and 2$\sigma$ regions shown in shading. 
This plot shows that the proportional approximation is consistent with the evolution in \textsc{Shark} in the most recent 8 Gyr, with deviations from this approximation occuring in the early Universe. 
We highlight that there are complexities in the nature of the modelling itself within \textsc{Shark} that make this comparison difficult. 
Instantaneous recycling of metals will likely cause the metallicity to build up more quickly than in reality, as an example. 
From the perspective of SED fitting, the age--metallicity degeneracy is greatest at this epoch where the discrepancy between the linear model and \textsc{Shark} is at its maximum. 
Furthermore, because the stellar populations formed early on in the history of a galaxy contribute to such a small fraction of the galaxy's observed flux, the constraining power of the SED fitting process is very small for this epoch. 
These factors make distinguishing between metallicity evolution models in the early Universe very difficult. 
Further work would be required with a larger range of semi-analytic and cosmological simulations to properly conclude the extent to which ``linearity" is physical, however such an analysis is well beyond the scope of this work.   

We reiterate that the implementation of the linear metallicity evolution is a significant improvement over models in which the metallicity is assumed to be constant over the duration of a galaxy's history. 
As discussed in Sec. \ref{sec:ProSpectFitting}, the linear metallicity evolution model is currently favourable over the closed-box model (presented in detail in \citealt{Bellstedt20b}), as the resulting metallicity distributions are more physical. 
This implies that the most recent epoch of evolution (to which SED fitting is most sensitive), is better modelled by linear metallicity evolution than the closed-box. 
\textsc{ProSpect} can be implemented with any model of metallicity evolution, and therefore this is straightforward to adapt in the future if desired.   

\section{Fitting the evolving plane}

\begin{figure}
	\centering
	\includegraphics[width=85mm]{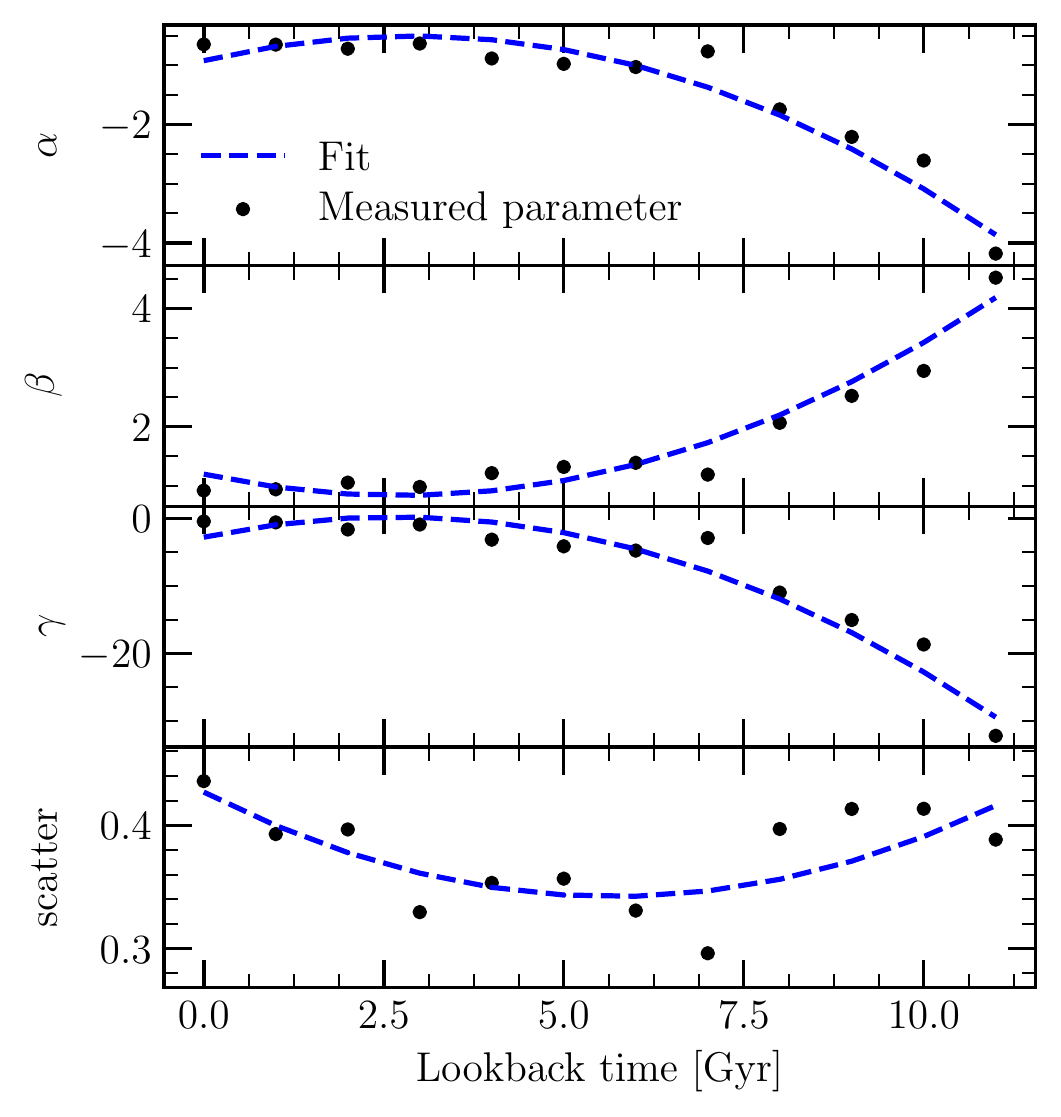}
	\caption{Measured parameters describing the fit to the plane in each epoch between 1 and 11 Gyr lookback times. The fit, showing the time evolution of the plane and presented in Equation \ref{eqn:Plane} is shown in blue.  }
	\label{fig:EvolvingPlaneParameters}
\end{figure}

The fit to the evolving plane parameters, as presented by Equation \ref{eqn:Plane} is presented in Fig. \ref{fig:EvolvingPlaneParameters}. Note that the evolution in the parameters is only significant beyond a lookback time of 7 Gyr. 

\end{document}